\documentclass[useAMS,usenatbib]{mn2e}
\usepackage{amssymb,amsmath,epsfig}

\begin{document}

\title[Detection of Disturbed Galaxy Morphologies]{New Image Statistics for Detecting Disturbed Galaxy Morphologies at High Redshift}
\author[P. E. Freeman et al.]{P.~E.~Freeman,$^1$\thanks{E-mail: pfreeman@cmu.edu} R.~Izbicki,$^1$ A.~B.~Lee,$^1$ J.~A.~Newman,$^2$ C.~J.~Conselice,$^3$ 
\newauthor A.~M.~Koekemoer,$^4$ J.~M.~Lotz$^4$ and M.~Mozena$^5$\\
$^1$Department of Statistics, Carnegie Mellon University, 5000 Forbes Avenue, Pittsburgh, PA 15213, USA\\
$^2$Department of Physics and Astronomy, University of Pittsburgh, 3941 O'Hara Street, Pittsburgh, PA 15260, USA\\
$^3$School of Physics and Astronomy, University of Nottingham, Nottingham NG7 2RD\\
$^4$Space Telescope Science Institute, 3700 San Martin Drive, Baltimore, MD 21218, USA\\
$^5$UCO/Lick Observatory and Department of Astronomy \& Astrophysics, University of California, Santa Cruz, CA 95064, USA
}

\date{5 June 2013}

\pagerange{\pageref{firstpage}--\pageref{lastpage}} \pubyear{2013}

\maketitle

\label{firstpage}

\begin{abstract}
Testing theories of hierarchical structure formation requires estimating the
distribution of galaxy morphologies and its
change with redshift.  One aspect of this investigation
involves identifying galaxies with disturbed morphologies (e.g., merging
galaxies).  This is often done by summarizing galaxy images using, e.g.,
the $CAS$ and Gini-$M_{20}$ statistics of \cite{Conselice03} and
Lotz et al.~(2004), respectively, and associating particular statistic values
with disturbance.  We introduce three statistics that
enhance detection of disturbed morphologies at high-redshift ($z \sim 2$):
the {\em multi-mode} ($M$), {\em intensity} ($I$), and {\em deviation}
($D$) statistics.  We show their effectiveness by training a 
machine-learning classifier, random forest, using
1,639 galaxies observed in the $H$-band by the 
{\it Hubble Space Telescope} WFC3, galaxies that had been 
previously classified by eye by
the CANDELS collaboration (\citealt{Grogin11}, \citealt{Koekemoer11}).
We find that the $MID$ statistics
(and the $A$ statistic of \citealt{Conselice03})
are the most useful for identifying disturbed morphologies.

We also explore whether human annotators are useful for identifying disturbed 
morphologies.  We demonstrate that they show limited ability to detect 
disturbance at high redshift, and that increasing their number
beyond $\approx$10 does not provably yield better classification performance.
We propose a simulation-based model-fitting algorithm 
that mitigates these issues by bypassing annotation.
\end{abstract}


\section{Introduction}

A thorough investigation of
cosmological theories of hierarchical structure formation
requires an accurate and precise estimate of the distribution of observed
galaxy morphologies and how it varies as a function of redshift.
One may attack this problem from a number of angles, including
determining the galaxy (major and minor) merger fraction at a range 
of redshifts.
Current estimates of the merger fraction 
at redshifts $z \leq$ 1.4 vary widely, from $\sim$ 0.01 to
$\sim$ 0.1 (see, e.g., \citealt{Lotz11} and references therein), with
quoted errors $\approx$0.01-0.03; at higher redshifts up to $z \sim$ 3,
merger fraction estimates rise to as high as 0.4 (e.g., \citealt{Conselice03b},
\citealt{Conselice08}, \citealt{Bluck12}).
Theory offers little guidance for resolving discrepancies among analyses:
while the dark matter halo-halo merger fraction has been
estimated consistently via simulations, uncertainty in the physical processes 
linking halos to underlying galaxies currently
precludes consistent estimation of the galaxy merger fraction
(e.g., \citealt{Bertone09}, \citealt{Jogee09}, \citealt{Hopkins10}).

As stated in, e.g., \cite{Lotz11},
post-merger morphologies are sufficiently ambiguous that we cannot use
local galaxies as an accurate tracer of the merger fraction and its evolution.
Even if all merger events did result in the formation of spheroidal
galaxies, the converse is not true:
not all spheroidal galaxies arise from mergers.
Thus to estimate the merger fraction and its evolution, we must seek out
ongoing mergers themselves.  
High-redshift merger sample sizes
range from the hundreds (e.g., \citealt{Lotz08},
\citealt{Jogee09}, \citealt{Kartaltepe10}) to $\approx$1,600 
(\citealt{Bridge10}), a regime in which human-based analysis 
(i.e., {\em annotation}) is feasible.  However, on-going
surveys such as the Cosmic Assembly Near-IR Deep Extragalactic Legacy Survey 
(CANDELS; \citealt{Grogin11}, \citealt{Koekemoer11}) will 
increase the number of putative mergers to the tens of thousands.

One approach to inferring merger activity involves detecting disturbed
morphologies within individual galaxies.  Ideally, they would manifest 
themselves either as
separate classes or as outliers within the evolving distribution of galaxy
shapes, a distribution that we would estimate directly from an 
adequately large set of galaxy images.
However, the direct use of galaxy images 
is both statistically and computationally intractable.
Thus we instead transform inherently high-dimensional (i.e., multi-pixel)
images into a lower-dimensional representations that retain
important morphological information, and we use them
to estimate discrete classes: early-type galaxies
vs.~late-type galaxies, mergers vs.~non-mergers, etc.  

Human annotators perform dimension reduction and
discretization implicitly (e.g., \citealt{Lintott08},
\citealt{Bridge10}, \citealt{Darg10}, \citealt{Kartaltepe10}), but
labeling galaxies is time consuming both in terms of infrastructure development
and implementation.  (Also, the inferential accuracy achieved by using
many non-expert annotators$-$i.e., by crowdsourcing$-$versus that achieved
by a smaller set of experts is an as-yet unresolved issue.)  
The main alternative to large-scale human labeling
is to extract low-dimensional summary statistics (or features)
from galaxy images, then to use the statistics and labels associated
with a small subset of images to train a computer-based classifier.
Of course, the
effectiveness of this approach hinges upon how well we actually 
retain important morphological information
when transforming image data, i.e., on whether we define an appropriate
set of statistics.

There are a myriad of statistics that one may extract from image data.
Common ones include the S\'ersic index and
bulge-to-disk ratio (found, e.g., using GALFIT; \citealt{Peng02}).
However, for our particular case of interest--detecting complex 
sub-structures in images of peculiar and irregular galaxies--statistics
that do not require model fitting are clearly optimal.
Such statistics include
the concentration, asymmetry, and clumpiness ($CAS$) statistics 
(e.g., \citealt{Bershady00} and \citealt{Conselice03}, hereafter C03),
and the Gini ($G$) and $M_{20}$
statistics (\citealt{Abraham03} and \citealt{Lotz04}, hereafter L04).\footnote{
We also note the multiplicity statistic $\Phi$ of \cite{Law07}, which we
do not apply in this work.}
Numerous authors apply these statistics; recent examples 
include \cite{Chen10}, \cite{Kutdemir10}, \cite{Lotz10a}, \cite{Lotz10b},
\cite{Conselice11}, \cite{Holwerda11}, \cite{Mendez11}, 
and \cite{Hoyos12}.  

Given this context,
there are three outstanding issues in galaxy morphology analysis that we
address in this work.
\begin{itemize}
\item The efficacy of the $CAS$ and $GM_{20}$ statistics for detecting
disturbed morphologies degrades
as galaxy signal-to-noise and size decrease, i.e., as redshift increases
(see, e.g., Figures 9 and 19 of \citealt{Conselice00},
Figures 5-6 of L04, and \citealt{Lisker08}).  In {\S}\ref{sect:feature}, 
we define
three new statistics (which we dub the $MID$ statistics) that
improve our ability to collectively detect peculiar and irregular galaxies
(which we dub {\em non-regulars}) as well as to detect merging\footnote{
Note that by ``merging," we mean ``merging and/or interacting."  We
downplay the explicit detection of interaction because we currently
only analyze each galaxy in isolation, without regard to
possible nearby galaxies, which clearly impedes our ability to detect 
interacting galaxies.}
systems themselves.  In {\S}\ref{sect:apply}, we apply these statistics
to the analysis of 1639 high-redshift ($z \sim$ 2) galaxies 
in the GOODS-South Early Release Science field (\citealt{Windhorst11}),
observed in the near-infrared regime by the 
Wide-Field Camera 3 (WFC3) on-board the {\it Hubble
Space Telescope} ({\it HST}).\footnote{
We note that statistics such as $C$ and $S$ were not developed for detecting
merging galaxies, per se.  However, as is shown in {\S}4,
including them in our analyses does not adversely affect the performance of our 
classification algorithm.
}
\item Authors generally apply the $CAS$ and $GM_{20}$ in a 
non-optimal manner, by projecting high-dimensional spaces
containing values for each observed galaxy
down to two-dimensional planes (e.g., $G$-$M_{20}$) and delineating
classes by eye.  In {\S}\ref{sect:anal}, we introduce 
the use of random
forest, a machine learning-based classifier that one can directly apply to
high-dimensional spaces of image statistics, to morphological analysis.
Ultimately, for analysis of large datasets, one would use random forest to
to train a classifier on a small subset of labeled galaxies, then apply it
to the unlabeled galaxies.
\item What is contribution of annotators and their biases to the overall error
budget in morphological analyses?  For instance, the fact that widely varying 
merger fraction estimates are generally associated with small error bars
indicates clearly the presence of (as yet unmodeled) systematic errors.
These errors (such as that associated with the inability of various authors
to agree on what defines a merger) may ultimately doom those 
morphological analyses in which humans play a role.  In {\S}\ref{sect:meta},
we first examine whether it is always beneficial to have more annotators 
look at each galaxy image, then ask whether it even necessary to invoke 
annotation if our ultimate goal is to constrain models of hierarchical
structure formation.
\end{itemize}
In {\S}\ref{sect:summary}, we summarize our results
and discuss future research directions.


\section{The MID statistics}

\label{sect:feature}

Let $f_{i,j}$ denote the observed flux at pixel $(i,j)$ of a given image
that has $n = n_x \times n_y$ pixels overall.  We assume that associated with
the image is a segmentation map that defines the extent of the object of
interest, such as is output by, e.g., the
source detection package {\tt Sextractor} (\citealt{Bertin96}).


\subsection{M(ulti-mode) statistic}

\label{sect:mstat}

Let $q_l$ denote an intensity quantile; for instance, $q_{0.8}$ denotes an
intensity value such that 80 percent of the pixel intensities inside the 
segmentation map are smaller than this value.
For a given value of $l$, we examine image pixels and define a new image
\begin{eqnarray}
g_{i,j} = \left\{ \begin{array}{cl} 1 & f_{i,j} \geq q_l \\ 0 & \mbox{otherwise} \end{array} \right. \,. \nonumber
\end{eqnarray}
This image will be mostly 0, with $m$ groups of contiguous pixels of value 1 
where the galaxy intensity is largest.  We determine the number of pixels
$A_{l,m}$ in each group, and order them in {\em descending} order (i.e., 
$A_{l,(1)}$ is the largest group of contiguous pixels for quantile $l$,
$A_{l,(2)}$ is the second-largest group, etc.).  We define the {\em area
ratio} for each quantile as
\begin{eqnarray}
R_l = \frac{A_{l,(2)}}{A_{l,(1)}} A_{l,(2)} \,, \label{eqn:ar}
\end{eqnarray}
This statistic is suited for detecting double nuclei within the segmentation
map, as the ratio $A_{l,(2)}/A_{l,(1)}$ tends towards 1 if 
double nuclei are present, and towards 0 if not.  Because this ratio
is sensitive to noise, we multiply it by $A_{l,(2)}$, which tends towards
0 if the second-largest group is a manifestation of noise.
The {\it multi-mode} statistic is the maximum $R_l$ value, i.e.,
\begin{eqnarray}
M = \max_l \, R_l \,. \label{eqn:mstat}
\end{eqnarray}

\begin{figure}
  \begin{minipage}[ht]{0.5\linewidth}
  \begin{center}
    \includegraphics[width=42mm]{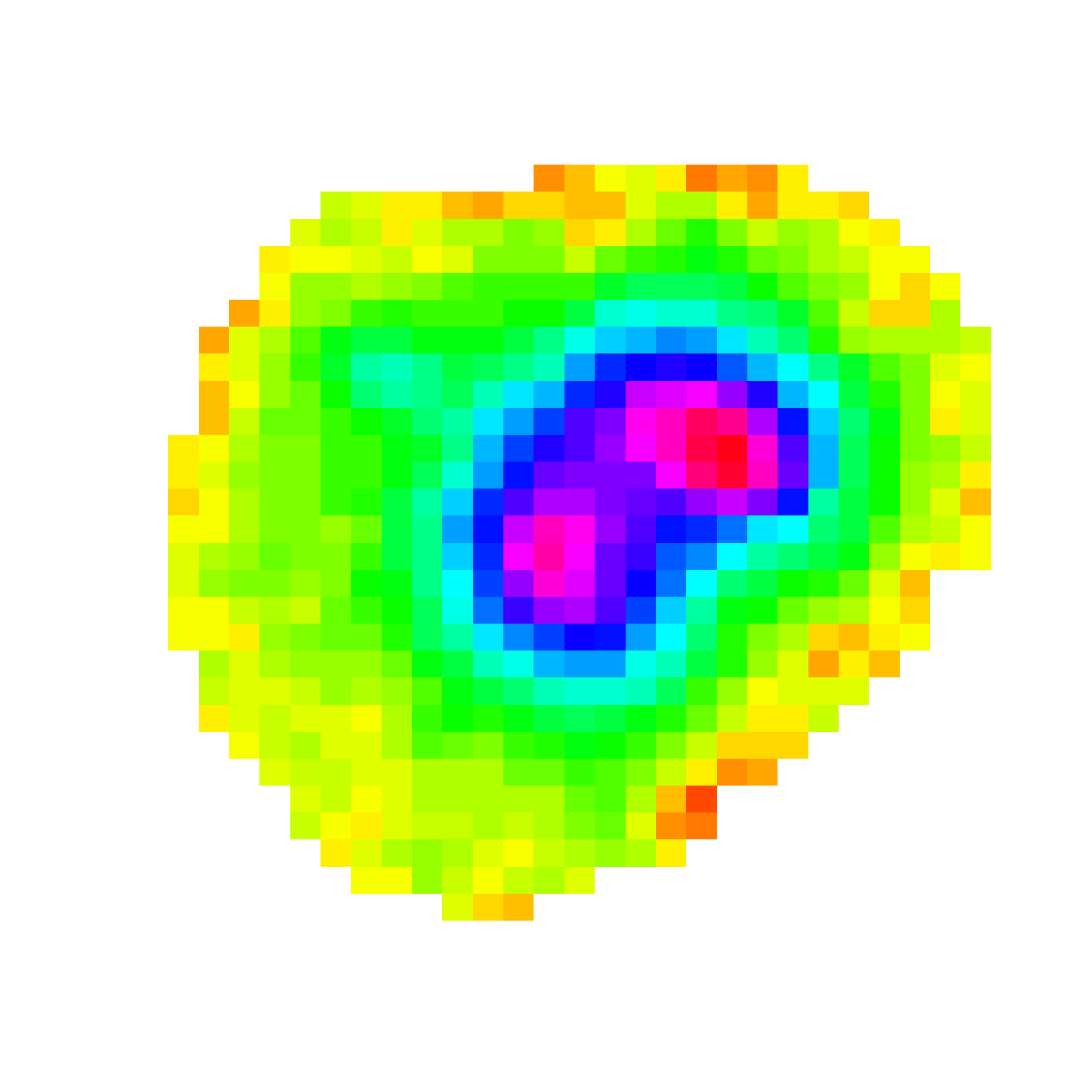}
  \end{center}
  \end{minipage}\hfill
  \begin{minipage}[ht]{0.5\linewidth}
  \begin{center}
    \includegraphics[width=42mm]{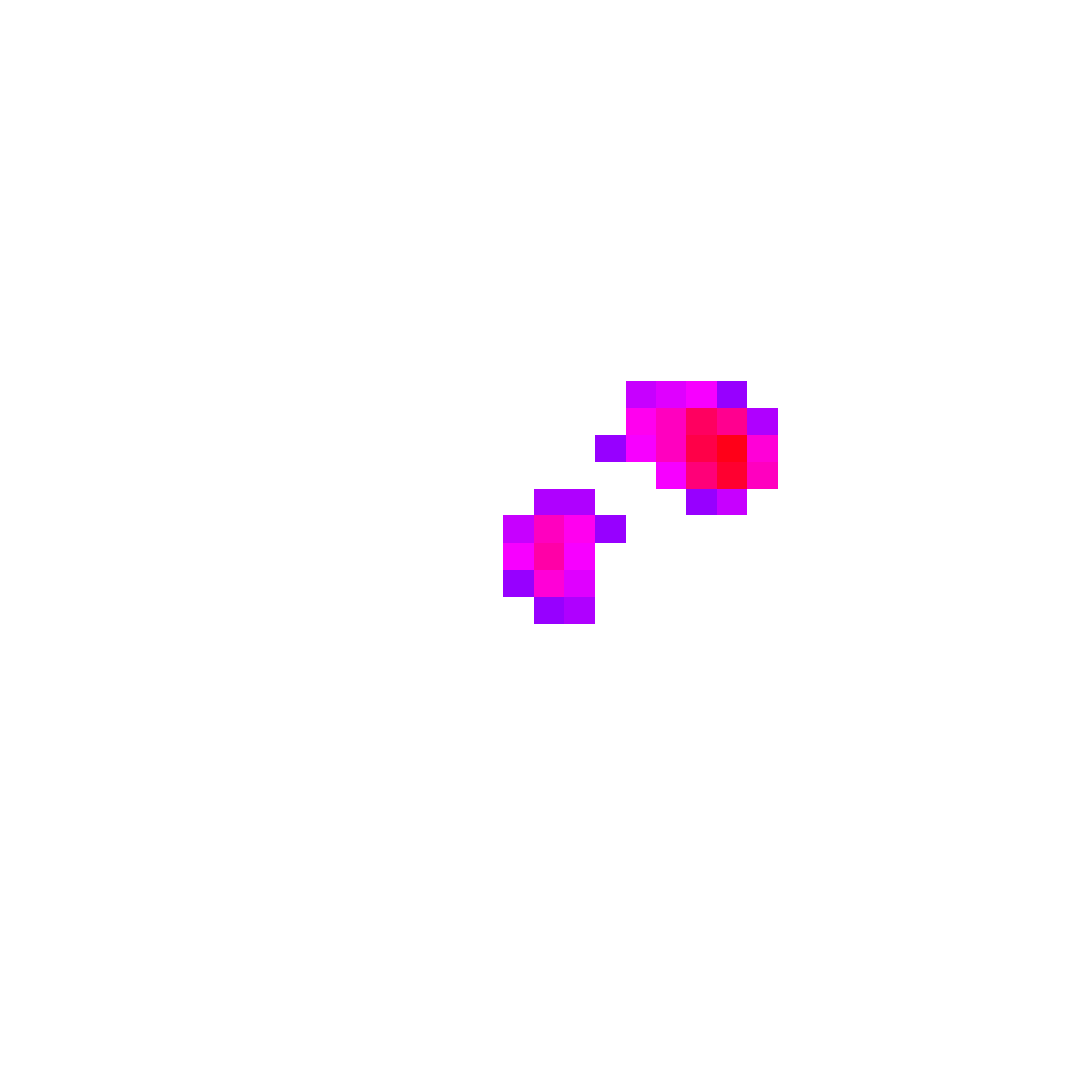}
  \end{center}
  \end{minipage}\hfill
\caption{Example of pixel grouping for computing the multi-mode
($M$) statistic.  To the left we display the $H$-band pixel intensities for
an example of a merger galaxy (see Figure \ref{fig:ERS2072}), while to
the right we show only those pixels associated with the largest
5.8\% of the sorted intensity values.  The two pixel regions have
areas $A_{(1)} = 21$ and $A_{(2)} = 14$, so that 
$R_{0.942} = (14/21)*14 = 9.33$.
The $M$ statistic is the largest of the $R$ values
computed for a sufficiently large number of threshold percentiles.
}
\label{fig:mstat}
\end{figure}

\subsection{I(ntensity) statistic}

\label{sect:istat}

The $M$ statistic is a function of the footprint areas of 
non-contiguous groups of image pixels, but does not take into account
pixel intensities.  To complement it,
we define a similar statistic, the {\em intensity} or $I$ statistic.
A readily apparent, simple definition of this statistic is the ratio of
the sums of intensities in the two pixel groups used to compute $M$.  However, 
this is not optimal, as in any given image it is possible, e.g., that a 
high-intensity pixel group with small footprint may not enter into the 
computation of $M$ in the first place.

\begin{figure}
  \begin{minipage}[ht]{0.5\linewidth}
  \begin{center}
    \includegraphics[width=42mm]{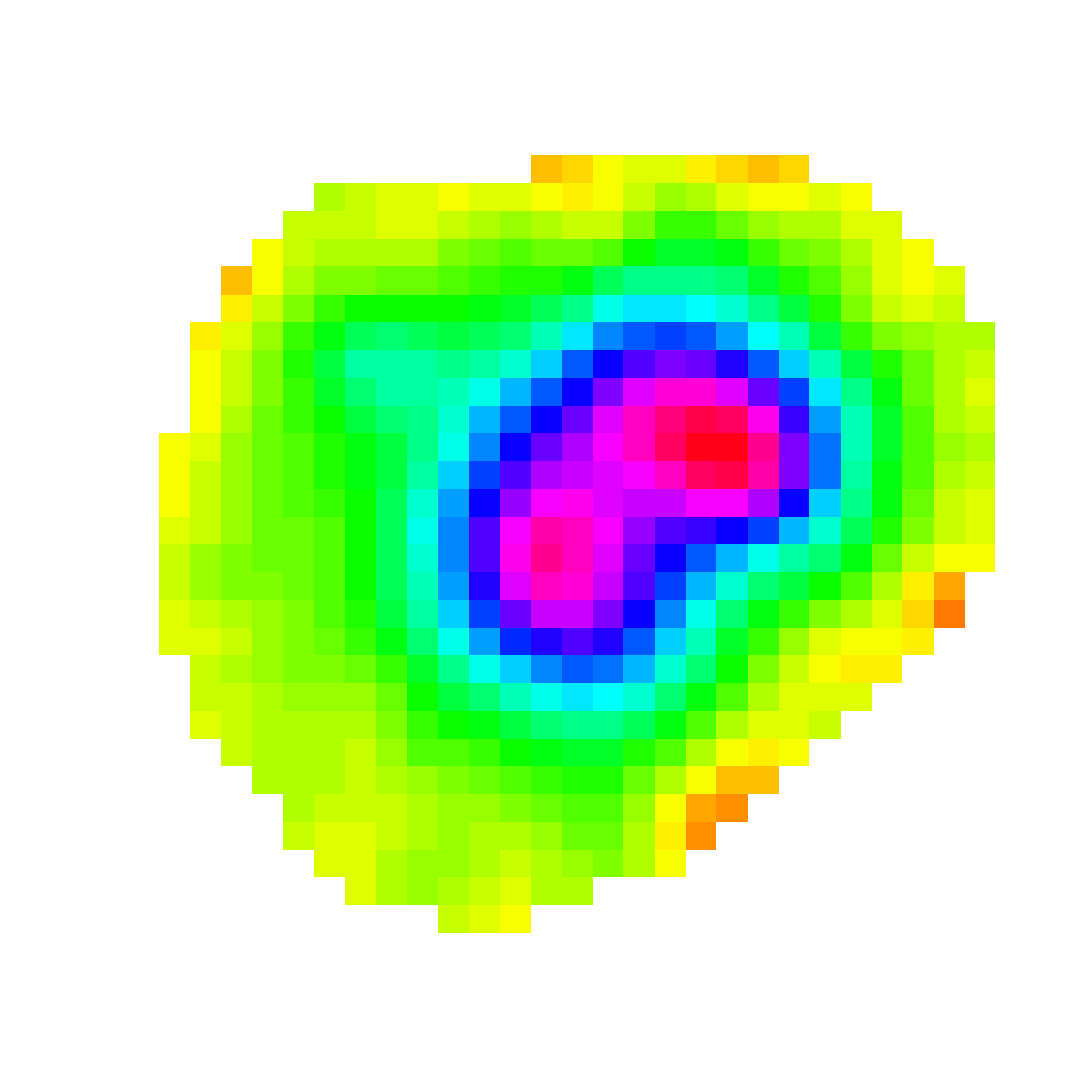}
  \end{center}
  \end{minipage}\hfill
  \begin{minipage}[ht]{0.5\linewidth}
  \begin{center}
    \includegraphics[width=42mm]{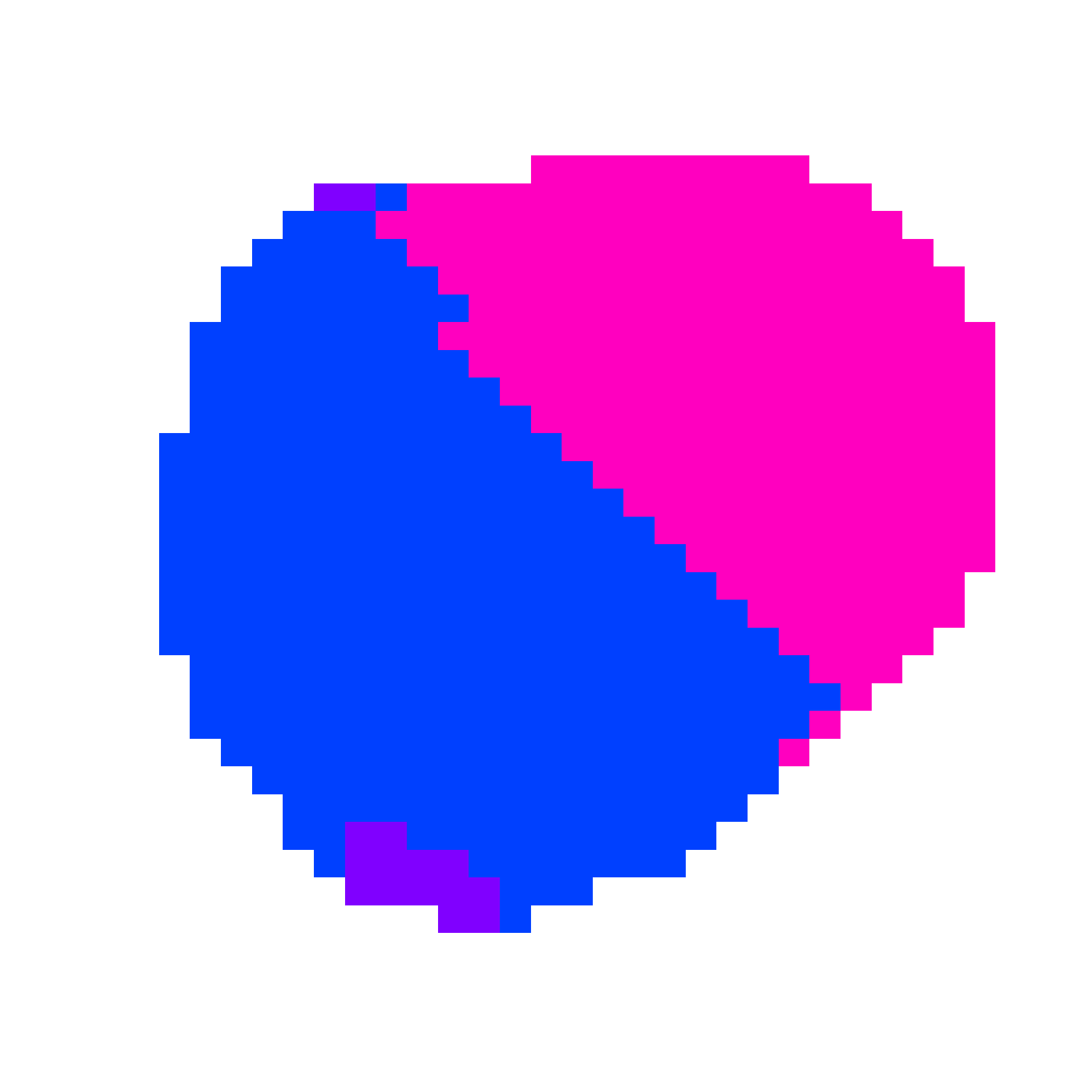}
  \end{center}
  \end{minipage}\hfill
\caption{Example of pixel grouping for computing the intensity
($I$) statistic.  To the left we display pixel intensities
for an example of a merger galaxy (see Figure \ref{fig:ERS2072}).
These data are smoothed using a symmetric Gaussian kernel of 
width $\sigma$ = 1 pixel, a sufficiently small scale to remove local 
intensity maxima caused by noise without removing local maxima intrinsic
to the galaxy itself.  (See the text for details on how we select the
appropriate smoothing scale $\sigma$.)
To the right we display pixel regions associated
with each local intensity maximum remaining after smoothing.
Pixel intensities
are summed within each region, with the intensity statistic then
being the ratio of the second-largest to largest sum.  In this example, 
the statistic is $I$ = 0.935.
}
\label{fig:istat}
\end{figure}

There are myriad ways in which one can define pixel groups over which 
to sum intensities.  In this work, we utilize a two-pronged approach.
First, we smooth the data in each image with a 
symmetric bivariate Gaussian kernel, 
selecting the optimal width $\sigma$ by maximizing the relative
importance of the $I$ statistic in correctly identifying morphologies
(i.e., how well we can differentiate classes using the $I$ statistic alone,
relative to how well we can differentiate classes by using other statistics
by themselves; see {\S}\ref{subsect:perf}).
Then we define groups using
maximum gradient paths.  For each pixel in the smoothed image, we 
examine the surrounding eight pixels and move to the one for which 
the increase in intensity is maximized, repeating
the process until we reach a local maximum.
A single group consists of all pixels linked to a 
particular local maximum.  (See Figure \ref{fig:istat}.)
Once we define all pixel groups,
we sum the intensities within each
and sort the summed intensities in descending order: $I_{(1)}$,
$I_{(2)}$,...  The intensity statistic is then
\begin{eqnarray}
I = \frac{I_{(2)}}{I_{(1)}} \,. \label{eqn:istat}
\end{eqnarray}

\subsection{D(eviation) statistic}

\label{sect:dstat}

Galaxies that are clearly irregular or peculiar will exhibit
marked deviations from elliptical symmetry.  A simple measure quantifying this
deviation is the distance from a galaxy's intensity centroid to the
local maximum associated with $I_{(1)}$, the pixel group with maximum
summed intensity.  We expect this quantity to cluster near zero for 
spheroidal and disk galaxies.  For those disk galaxies with
well-defined bars and/or spiral arms, we would still expect near-zero
values, as between the bulge and generally expected structure symmetry,
both the intensity centroid and the maximum associated with $I_{(1)}$ 
should lie at the galaxy's core.  

We define the intensity centroid of a galaxy as
\begin{eqnarray}
(x_{\rm cen},y_{\rm cen}) = \left( \frac{1}{n_{\rm seg}} \sum_i \sum_j i f_{i,j} , \frac{1}{n_{\rm seg}} \sum_i \sum_j j f_{i,j} \right) \,. \nonumber 
\end{eqnarray}
with the summation being over
all $n_{\rm seg}$ pixels with the segmentation map.
The distance from $(x_{\rm cen},y_{\rm cen})$ to the maximum associated
with $I_{(1)}$ will be
affected by the absolute size of the galaxy; it generally will be larger 
for, e.g., galaxies at lower redshifts.  Thus we normalize the distance
using an approximate galaxy ``radius," $\sqrt{n_{\rm seg}/\pi}$.
The {\it deviation} statistic is then
\begin{eqnarray}
D = \sqrt{\frac{\pi}{n_{\rm seg}}} \sqrt{(x_{\rm cen}-x_{I_{(1)}})^2 + (y_{\rm cen}-y_{I_{(1)}})^2} \,. \label{eqn:dstat}
\end{eqnarray}

We have designed the $D$ statistic to capture evidence of galaxy asymmetry.
It is thus complimentary to the $A$ statistic defined in C03,
which is computed
by rotating an image by 180$^{\circ}$, taking the absolute difference between
rotated and unrotated images, normalizing by the value of the unrotated image,
and summing the resulting values pixel-by-pixel.  In {\S}\ref{sect:apply}
we compute both $A$ and $D$ for a sample of high-redshift galaxies and
show that while there is a large positive sample correlation coefficient between
the two statistics, there are many
instances where $D$ captures stronger evidence of asymmetry than $A$, and
vice-versa, demonstrating that $D$ and $A$ are not simply redundant.

\section{Statistical analysis: random forest}

\label{sect:anal}

We use the $MID$ (and other) statistics to populate an $p$-dimensional
space, where each axis represents the values of the $i^{\rm th}$ statistic
and each data point represents one galaxy.  Ideally, in this space,
the set of points
representing, e.g., visually identified mergers is offset from those 
representing non-mergers.  To determine an optimal boundary between these
point sets directly in the $n$-dimensional space, we apply 
machine-learning based methods of regression and classification.
To ensure robust results, it is 
good practice to apply a number of methods to see if any one or more
provide significantly better results.  In our analysis of {\it HST} data
in {\S}\ref{sect:apply}, we tested four algorithms:
random forest, lasso regression, support vector machines, and principal
components regression (for details on these, see, e.g., HFT09).
We found that the results of applying each were similar: in the vast majority 
of cases, galaxies were either classified correctly or incorrectly by all 
four algorithms.  Thus in this work we describe only the conceptually simplest 
of the four algorithms, random forest (\citealt{Breiman01}).  For an example of
an analysis code that uses random forest, see Appendix \ref{app:ranfor}.

\subsection{Random Forest Regression}

The first step in applying random forest is a regression step: 
we randomly sample 50\% of the galaxies
(i.e., populate a training set)\footnote{
The size of the training set is arbitrary.  Larger training sets 
generally yield better final results.  In this work, our principal goal
is to demonstrate the efficacy of the $MID$ statistics relative to
other, commonly used ones, and so we do not explicitly address the issue
of optimizing the training set size.}
and regress the fraction of annotators,
$Y_i \in [0,1]$, who view galaxy $i$ as a non-regular/merger
upon that galaxy's set of image statistics.
In random forest,
bootstrap samples of the training set are used to
grow $t$ trees (e.g., $t$ = 500), each of which has $n$
nodes (e.g., $n$ = 8).  At each node in each tree, a random subset of 
size $m$ of the statistics is chosen (e.g., $C$, $M_{20}$,
and $I$ may be chosen from the full set of statistics).  
The best split of the data along
each of the $m$ axes is determined, with the one overall best split
retained.  This process is repeated for each node and each tree;
subsequently the training data are pushed down each tree to determine
$t$ class predictions for each galaxy
(i.e., to generate a set of $t$ numbers for each galaxy, 
all of which are either 0 or 1).
Let $s_i \in [0,t]$ equal the sum of the class predictions
for galaxy $i$; then the random forest prediction for the
galaxy's classification is ${\hat Y}_i = s_i/t$.

While fitting the training data, random forest keeps track of all $t$
trees it creates, i.e., all of the data splits it performs.  Thus any
new datum (from the remaining 50\% of the data, or the test set) may be
``pushed" down these trees, resulting again in $t$ class predictions
and thus a predicted response (or dependent) variable ${\hat Y}_i = s_i/t$.

\subsection{Random Forest Classification}

Once we predict response variables for the test set galaxies, we perform
the second step of random forest, the classification step.  First, the
continuous fractions $Y_i$ associated with the test set galaxies are
mapped to $Y_{{\rm class},i} = 0$ if $Y_i < 0.5$ and $Y_{{\rm class},i} = 1$
if $Y_i > 0.5$.  (Ties are broken by randomly assigning galaxies to classes.)
Then the predicted responses for the test set galaxies, ${\hat Y}_i$,
are also mapped to discrete classes.  Intuitively, one might expect the 
splitting point between predicted classes, $c$, to be 0.5.  However,
because the proportions of regular and non-regular galaxies in the training
set are unequal,
as are the proportions of non-merger and merger galaxies,
regression will be biased towards fitting the galaxies of the
more numerous type well.  For instance, regular galaxies outnumber non-regular
galaxies by approximately three-to-one, so {\it a priori} we expect 
the best value for $c$ to be around 1/(3+1) = 0.25.
To determine $c$, we select a sequence of values $\{c_1,c_2,\cdots,c_n\}
\in [0,1]$, and for each $c_j$
we map the predicted responses to two classes, e.g., we call all
galaxies with predicted responses $\hat{Y}_i > c_j$ non-regulars/mergers.
Call this classifying function $h(X,c_j)$, where $X$ is the set of observed
statistics.  Our estimate of risk as a function
of $c_j$ is the the overall proportion of misclassified galaxies:
\begin{eqnarray}
\hat{R}_j &=& \hat{P}[h(X,c_j) = 0 \vert Y = 1] + \hat{P}[h(X,c_j) = 1 \vert Y = 0] \nonumber
\end{eqnarray}
where $\hat{P}[h(X,c_j) = 0 \vert Y = 1]$ and
$\hat{P}[h(X,c_j) = 1 \vert Y = 0]$ are the estimated probabilities of
misclassifying a non-regular/merger and a regular/non-merger, respectively.
We seek the minimum value for this estimate of risk.
We smooth the discrete function $\hat{R}_j = f(c_j)$ with a Gaussian profile
of width 0.05, and choose as our final value of $\hat{c}$
that value for which the smoothed function $\hat{R}(c)$ is minimized.\footnote{
We note that this algorithm produces similar results to the
Bayes classifier (see, e.g., Chapter 2 of HFT09), 
which sets $c = l_0/(l_0+l_1)$.}

\subsection{Measures of Classifier Performance}

\label{subsect:perf}

\begin{table}
\caption{Confusion Matrix: Definitions}
\begin{tabular}{rcc}\hline
 & Predicted & Predicted \\
 & Regular/ & Non-Regular/ \\
 & Non-Merger & Merger \\
\hline
Actual & TN & FP \\
Regular/ & (true negatives) & (false positives) \\
Non-Merger \\
\\
Actual & FN & TP \\
Non-Regular/ & (false negatives) & (true positives) \\
Merger \\
\hline
\end{tabular}
\label{tab:confmat}
\end{table}

We use a number of measures of classifier performance.
\begin{itemize}
\item {\em Sensitivity.} The proportion of non-regular/merger galaxies that
are correctly classified: $TP/(TP+FN)$.  (This is also dubbed completeness.)
\item {\em Specificity.} The proportion of regular/non-merger galaxies that are
correctly classified: $TN/(TN+FP)$.
\item {\em Estimated Risk.} The sum of 1 $-$ sensitivity and 1 $-$ specificity.
\item {\em Total Error.} The proportion of misclassified galaxies.
\item {\em Positive Predictive Value (PPV).} The proportion of actual
non-regular/merger galaxies among those predicted to be non-regulars or 
mergers: $TP/(TP+FP)$.  (This is also dubbed purity.)
\item {\em Negative Predictive Value (NPV).} The proportion of actual
regular/non-merger galaxies among those predicted to be regulars or 
non-mergers: $TN/(TN+FN)$.
\end{itemize}
We define the symbols used above in Table \ref{tab:confmat}.

Random forest assesses the efficacy of each statistic 
for disambiguating classes by computing Gini importance scores
for each (see, e.g., Chapter 9 of HFT09; note that the Gini importance
score differs from the Gini statistic $G$).
At any given node of any tree, the $n$ samples to be split belong to two
classes (e.g., merger/non-merger), with proportions $p_1 = n_1/n$ and
$p_2 = n_2/n$.  A metric of class impurity at this node is
$i = 1 - p_1^2 - p_2^2$.  The samples are then split along the axis associated
with one chosen image statistic, with proportions
$p_l = n_l/n$ and $p_r = n_r/n$ being assigned to two daughter nodes.  New
values of the impurity metric are computed at each daughter node; call 
these values $i_l$ and $i_r$.  The reduction in impurity 
achieved by splitting the data is
$\Delta i = i - p_l i_l - p_r i_r$.  
Each value of $\Delta i$ is associated with one image statistic; the average
of the $\Delta i$'s for each image statistic over all nodes and 
all trees is the Gini importance score.

Note that in this work, we are not 
as concerned with the absolute importance score of each statistic (which is
not readily interpretable) as we are with relative scores derived by, e.g.,
dividing importance scores by the maximum observed importance score value.
Relative scores are sufficient to allow us to rank the image
statistics in order of how useful they are by themselves in classification.
They also allow us
to reduce the dimensionality of our statistic space, if necessary,
by eliminating those that are not as useful for disambiguating
classes.  This point is not important in the context of implementing random
forest$-$the random forest algorithm is computationally efficient even when
presented with very high dimensional spaces of statistics$-$but does become
important if we are to implement density-estimation-based analysis schemes
like that discussed in {\S}\ref{sect:meta_less}.

\section{Application to HST images of GOODS-S galaxies}

\label{sect:apply}

%

\begin{figure}
\includegraphics[width=84mm]{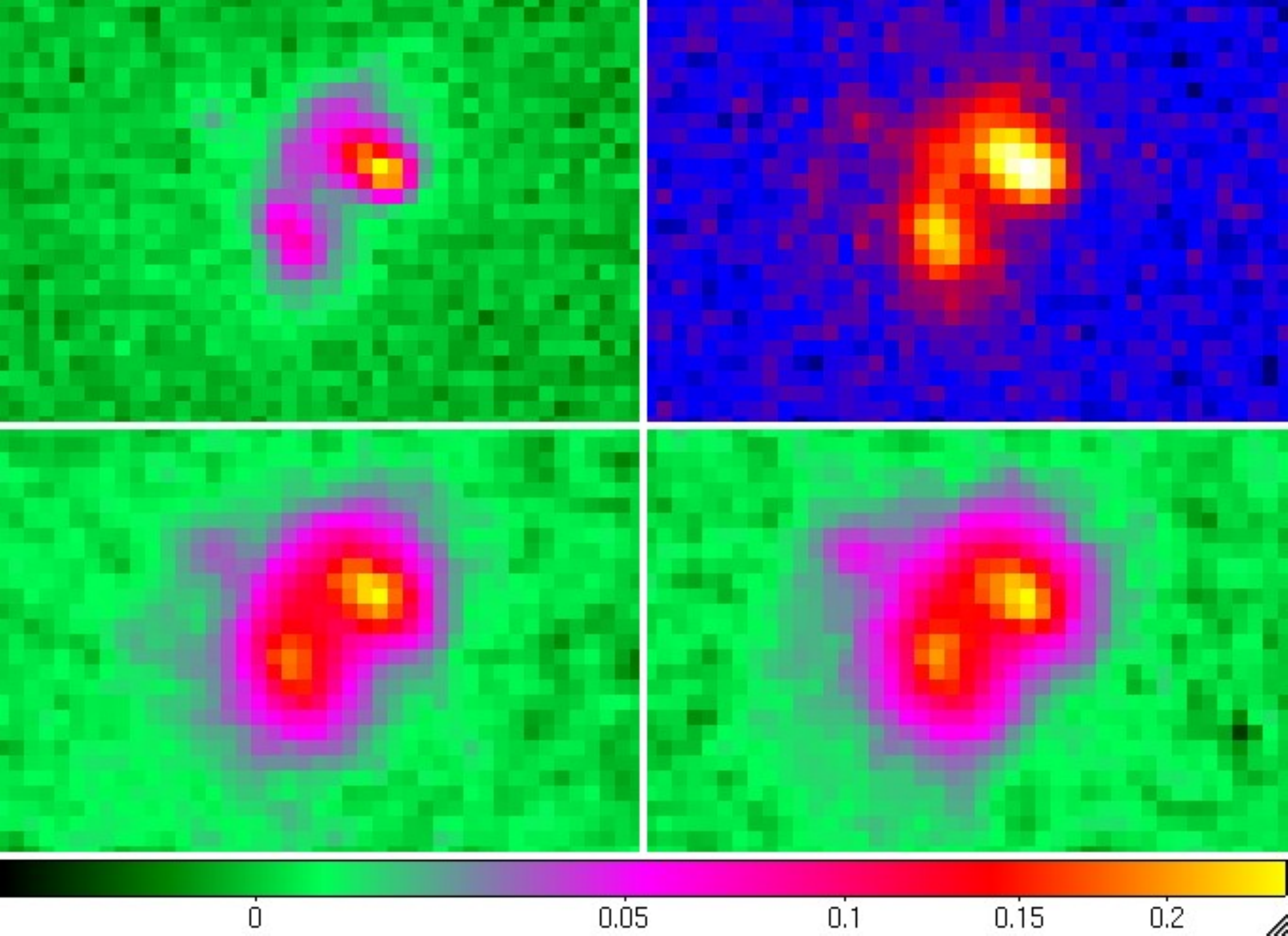}
\caption{Example of a galaxy identified as a merging galaxy by
CANDELS team annotators.  
{\it Top Left and Right:} 
{\it HST} ACS $V$- and $z$-band images, respectively.
{\it Bottom Left and Right:}
{\it HST} WFC3 
$J$- and $H$-band images, respectively.  Note the clear presence of
two nuclei in the $J$-band image.
}
\label{fig:ERS2072}
\end{figure}

We demonstrate the efficacy of the $MID$ statistics for detecting
non-regular and merging galaxies by analyzing
$H$- and $J$-band {\it Hubble Space Telescope} WFC3 images
of the northernmost part of the
Great Observatories Origins Deep Survey (GOODS) South field
(the Early Release Science fields; see \citealt{Windhorst11} and 
references therein).  
These images have been analyzed by members of the Cosmic Assembly
Near-IR Deep Extragalactic Legacy Survey (CANDELS; \citealt{Grogin11},
\citealt{Koekemoer11}) team.
They ran {\tt SExtractor}
to extract a catalog of 6178 putative sources and associated segmentation maps
in the $H$-band images,
with the input parameters set so as to optimize the detection and deblending
of galaxies at $z \sim 2$ (D.~Kocevski, private communication).
Subsequent visual morphological evaluation of these sources by, typically,
three or four CANDELS team members
yielded a set of 1639 galaxies with isophotal magnitudes $H < 25$
(Kartaltepe et al.~2012, in preparation).
In Figure \ref{fig:ERS2072} we show an example of one galaxy from the catalog
that annotators identified as undergoing a merger.

CANDELS-team labels
are based primarily on $H$-band images, with data in the other three
bands used to inform labeling.  An annotator may first
choose one (or more) ``main" morphological class(es) (e.g., spheroid,
disk, irregular), then indicate whether the galaxy is in the process
of merging (or interacting within or beyond the segmentation map).
Determining the fraction of annotators identifying a particular galaxy as
a merger is straightforward.  However,
because individual annotators could register more than one vote, 
and because only the final vote totals are available, it is impossible to
determine how many annotators labeled any one galaxy both as
irregular {\em and} a merger as opposed to selecting only one of the two
(as well as to determine the number of annotators, period).
We thus estimate the fraction of votes for non-regularity,
$f = V_\ast/V$, where $V$ is the total number of votes,
by defining $V_\ast$ as
\begin{eqnarray}
V_\ast \equiv {\rm mean}(V_\ast^{\rm low},V_\ast^{\rm high}) \,, \nonumber
\end{eqnarray}
with 
\begin{eqnarray}
V_\ast^{\rm low} &=& {\rm max}\{ {\rm votes\,for\,irregular,votes\,for\,merger} \} \nonumber
\end{eqnarray}
and
\begin{eqnarray}
V_\ast^{\rm high} &=& {\rm max}\{ {\rm votes\,for\,irregular\,+\,votes\,for\,merger}, V \} \nonumber
\end{eqnarray}
representing lower and upper bounds on the number of annotators that voted
for merger {\em or} irregular.

\subsection{Base Analysis: H-band Data}

\label{subsect:hband}

In our base analysis, we apply random forest regression and classification
to detect non-regulars and mergers within a labeled set
of 1639 galaxies observed in the $H$-band.
For each galaxy, we have as the predictor (or independent) variables
\begin{itemize}
\item the $MID$ statistics, which we compute using postage stamp images 
(generally 84 $\times$ 84 pixels) and a segmentation map; and 
\item the $CAS$ (a la C03) and
$GM_{20}$ statistics (a la L04), as provided by the 
CANDELS collaboration.
\end{itemize}
Recall that prior to computing the $I$ and $D$ statistics,
we smooth the data with a symmetric Gaussian kernel of width $\sigma$ 
to mitigate the effect of local intensity maxima caused by noise (see
{\S}\ref{sect:istat}).
We choose $\sigma$ by maximizing the importance of the $I$ statistic;
here, $\sigma$ = 1 pixel for both the non-regular and merger analyses.
The continuous response variable $Y \in [0,1]$ is
\begin{itemize}
\item $f = V_\ast/V$, as estimated from visual annotations by members
of the CANDELS collaboration.  The numbers of galaxies with vote fractions
favoring non-regularity of
$f > 0.5$ and $f = 0.5$ are 337 and 163, respectively; the analogous numbers
for the merger analysis are 109 and 71.
\end{itemize}

\begin{figure}
\includegraphics[width=84mm]{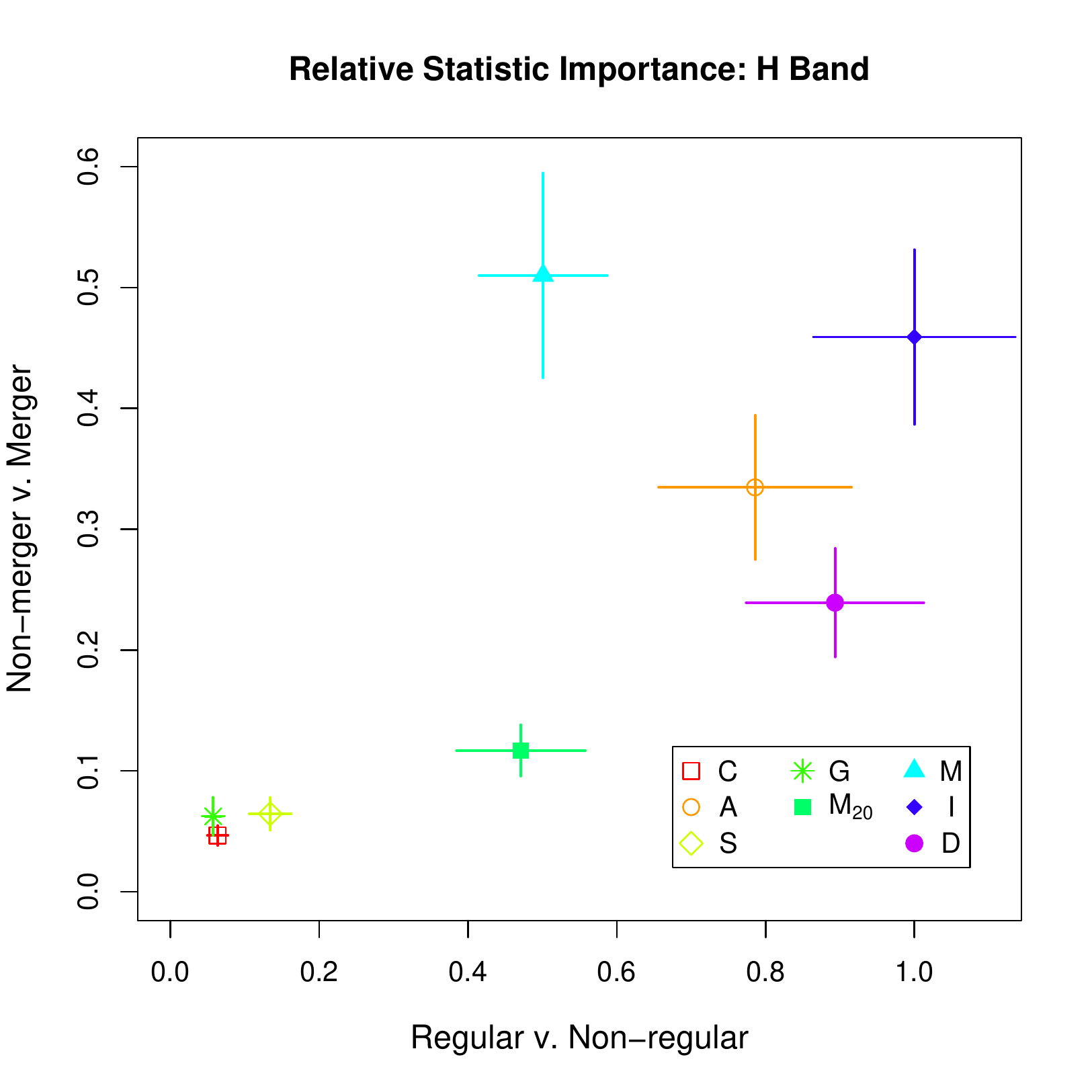}
\caption{Relative importance of statistics in differentiating between
regular and non-regular galaxies (x-axis) and non-merger and merging galaxies
(y-axis) in CANDELS-team-processed $H$-band data, as output by the random
forest classification algorithm.  The values of all data 
points have been normalized
by the value of the $I$ statistic for the regular/non-regular analysis.
See {\S}\ref{subsect:perf} for the definition of statistic importance, and
{\S}\ref{subsect:hband} for practical interpretation of the results.
The error bars indicate sample standard deviation given 1,000 separate
runs of random forest, and thus are not measures
of the standard error of the mean; the 1$\sigma$ uncertainties in
the means are given by shrinking the error bars by a factor of $\sqrt{1000}$
= 31.62.
}
\label{fig:Himp}
\end{figure}

\begin{figure}
\includegraphics[width=84mm]{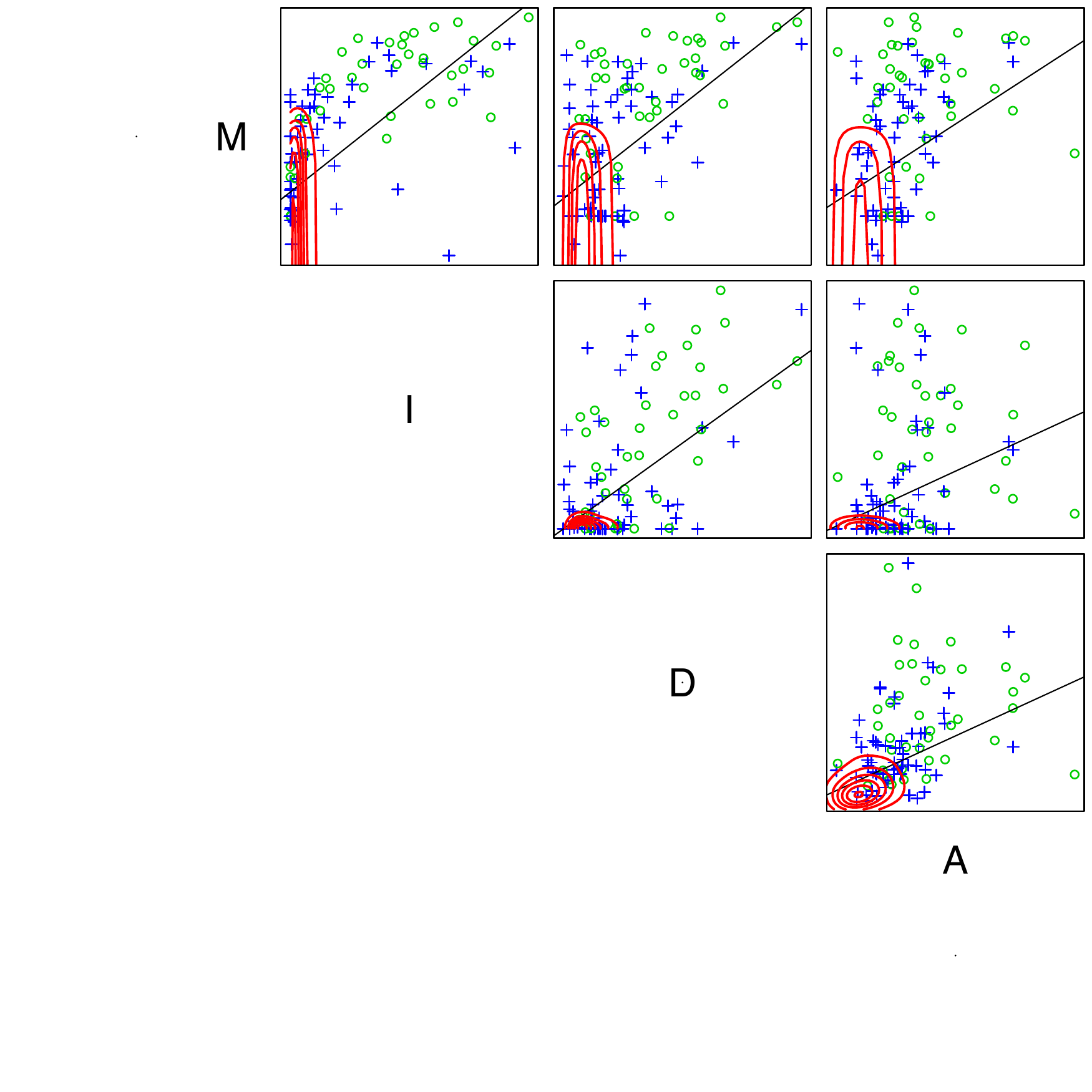}
\caption{Scatter plots of the $MID$ and $A$ statistics as computed for
CANDELS-team-processed $H$-band images.
Green circles represent galaxies visually identified as
mergers and blue crosses represent non-regulars that are not also mergers.
The red lines are contours indicating the density of regular galaxies.
The non-zero slopes of the black line, the best-fit linear 
regression functions, indicates the expected positive correlations between 
each of these statistics.
Note that for increased clarity, only 100 randomly selected 
non-regulars/mergers are displayed.
}
\label{fig:cor}
\end{figure}

In Figure \ref{fig:Himp}, we display the relative importance of each of the
$CAS$-$GM_{20}$-$MID$ statistics for detecting non-regulars (x-axis)
and mergers (y-axis).
These values are normalized such that the value of the most important
statistic, the $I$ statistic in the regular/non-regular analysis, is one.
(To interpret relative importance, note
for instance that the $M$ statistic by itself performs about half as well at
differentiating regulars from non-regulars as the $I$ statistic by itself.
This implies that there is greater separation between the distributions of $I$
statistics observed for regulars and for non-regulars than there is for the
analogous $M$ statistic distributions.  See Figure \ref{fig:cor}.)
The error bars in Figure \ref{fig:Himp} represent the standard deviations of
the populations from which we sample importance values, and are estimated
by running random forest 1,000 times, i.e., by
randomly dividing the 1,639 galaxies into training and test sets 1,000
times and recording variable importance and measures of classifier 
performance each time.  
Several conclusions are readily 
apparent when we examine Figure \ref{fig:Himp}:
(a) the $I$ statistic is the most important for detecting
both non-regulars and mergers; (b) as expected, the $M$ statistic 
outperforms the $D$ statistic in merger detection, and vice-versa for 
detection of non-regulars; and (c) the $MID$ statistics, along with the
$A$ (asymmetry) statistic of C03, are far more important
for identifying non-regulars/mergers than the other statistics we examine.  

Figure \ref{fig:cor} displays projections of the four-dimensional space 
of image statistics 
defined by the $MID$ and $A$ statistics and indicates the 
distributions of these statistics for
mergers (green points), non-regulars that are 
not mergers (blue points), and regulars (red contour lines).
The reader should intuitively picture classification via random forest as 
placing vertical and horizontal lines on these plots so as to maximize
the proportion of, e.g., non-regulars on one side and the proportion of
regulars on the other.  Given this intuitive picture, the 
relative efficacy of, e.g., the $I$ statistic with respect to the
other statistics is clearly evident.
Also evident from Figure \ref{fig:cor} is our relative inability to
separate mergers and irregular galaxies using $MID$ and $A$ alone.
Further work is needed to develop image statistics that will
optimize merger-irregular class separation.

The results in Figure \ref{fig:Himp} were generated by analyzing the
entire eight-dimensional space of image statistics with random forest.  
Given that
the number of possibly useful statistics will only increase in the future, it
is important to determine we can disregard any of our current statistics
with little, if any, loss in classifier performance.  
(This is not a trivial issue, as we discuss in {\S}\ref{sect:meta_less}:
the number of statistics we incorporate may limit future analyses.)
To that end, we define and analyze
three reduced statistic sets for both the regular/non-regular case
($ID$,$A$-$ID$,$A$-$MID$) and the non-merger/merger case
($MI$,$A$-$MI$,$A$-$MID$), based on rankings of relative statistic
importance.

See Tables \ref{tab:HRNR} and \ref{tab:HNMM}.
The interpretation of these tables depends on the performance metric
one prefers: e.g., sensitivity (or
catalog completeness), estimated risk, or PPV (or catalog purity), etc.
If we assume that one would wish to 
strike a balance between all three of these measures, we find that
the reduced statistic set $A$-$ID$ is sufficient for disambiguating 
regulars from non-regulars, while one requires the additional information
carried by the full set of statistics to disambiguate mergers from non-mergers.

\begin{table}
\caption{Classifier Performance\,$\times$\,10$^4$\,$-$\,$H$-band/Non-Regular}
\begin{tabular}{ccccc}
\hline
 & $ID$ & $A$-$ID$ & $A$-$MID$ & Full \\
\hline
sens   & 7562 $\pm$ 14 & 7907 $\pm$ 11 & 7873 $\pm$ 11 & 7874 $\pm$ 13 \\
spec   & 8071 $\pm$ 13 & 8187 $\pm$ 7  & 8136 $\pm$ 7  & 8181 $\pm$ 10 \\
risk   & 4366 $\pm$ 8  & 3906 $\pm$ 8  & 3991 $\pm$ 8  & 3945 $\pm$ 8 \\
toterr & 2056 $\pm$ 7  & 1883 $\pm$ 4  & 1930 $\pm$ 4  & 1897 $\pm$ 5  \\
PPV    & 5712 $\pm$ 13 & 5943 $\pm$ 9  & 5865 $\pm$ 9  & 5977 $\pm$ 11 \\
NPV    & 9089 $\pm$ 4  & 9215 $\pm$ 4  & 9199 $\pm$ 4  & 9194 $\pm$ 4  \\
\hline
\label{tab:HRNR}
\end{tabular}
This table displays sample mean 
plus/minus 1$\sigma$ standard error, 
each multipled by 10$^4$.
\end{table}

\begin{table}
\caption{Classifier Performance\,$\times$\,10$^4$\,$-$\,$H$-band/Merger}
\begin{tabular}{ccccc}
\hline
 & $MI$ & $A$-$MI$ & $A$-$MID$ & Full \\
\hline
sens   & 6917 $\pm$ 22 & 7520 $\pm$ 22 & 7790 $\pm$ 17 & 7816 $\pm$ 17 \\
spec   & 8683 $\pm$ 14 & 8546 $\pm$ 15 & 8564 $\pm$ 9 & 8591 $\pm$ 8 \\
risk   & 4400 $\pm$ 13 & 3934 $\pm$ 13 & 3646 $\pm$ 13 & 3594 $\pm$ 13 \\
toterr & 1477 $\pm$ 12 & 1547 $\pm$ 12 & 1506 $\pm$ 7  & 1480 $\pm$ 6  \\
PPV    & 3562 $\pm$ 20 & 3525 $\pm$ 20 & 3541 $\pm$ 13 & 3603 $\pm$ 13 \\
NPV    & 9662 $\pm$ 2  & 9723 $\pm$ 2  & 9752 $\pm$ 2  & 9753 $\pm$ 2  \\
\hline
\label{tab:HNMM}
\end{tabular}
This table displays sample mean 
plus/minus 1$\sigma$ standard error, 
each multipled by 10$^4$.
\end{table}

\subsection{Effect of changing the observation wavelength: J-band data}

\label{subsect:jband}

Recall that CANDELS team labels are based primarily on how
galaxies appear in the $H$ band.  However, in order to, e.g.,
differentiate true mergers from galaxies exhibiting disk instabilities,
we will need to extend the application of our statistics to other
wavelength regimes.  This extension is the subject of a future work;
here, we make a preliminary assessment of the robustness of the $MID$
statistics as a function of wavelength by applying them to the $J$-band
images associated with our galaxy sample.

\begin{figure}
\includegraphics[width=84mm]{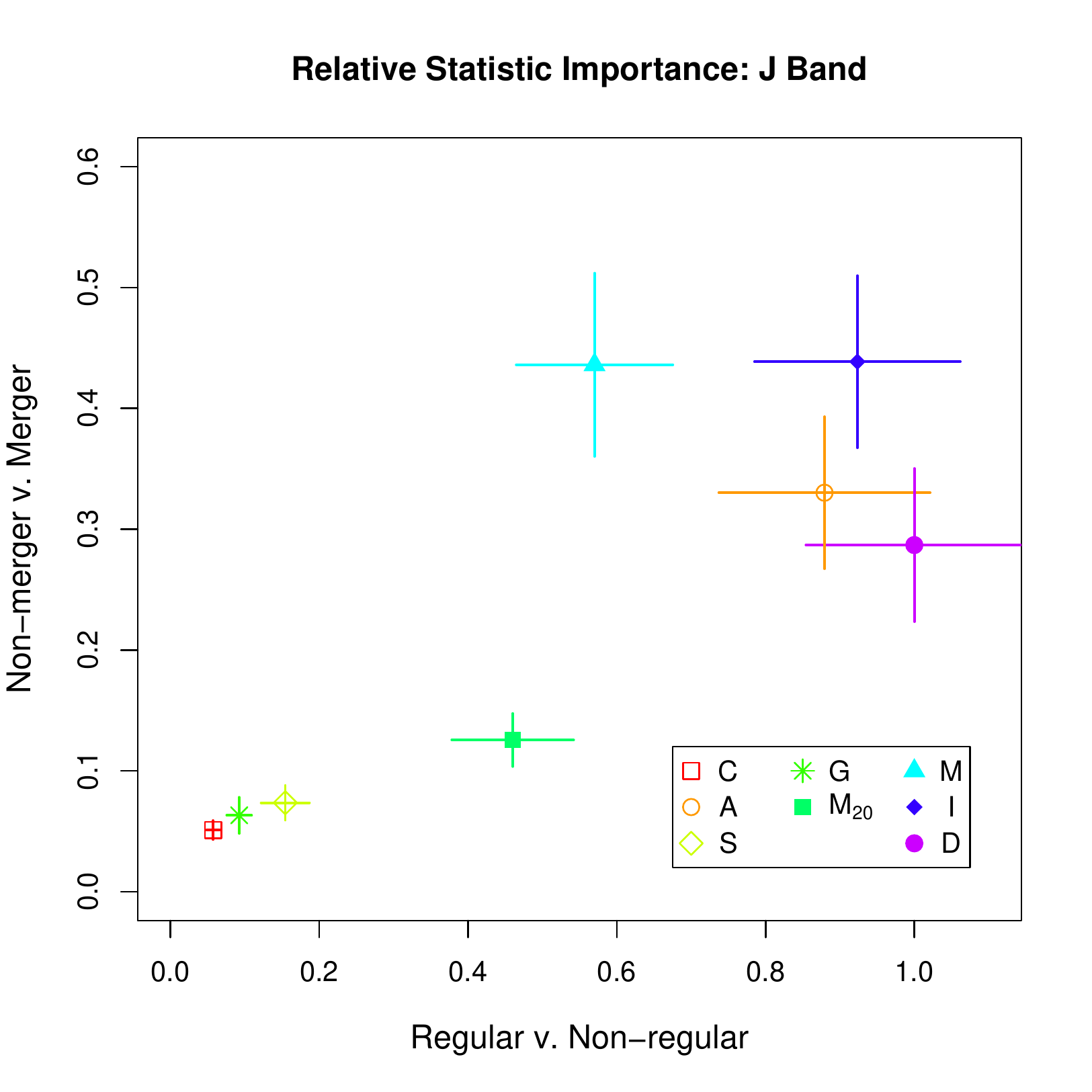}
\caption{Same as Figure \ref{fig:Himp}, but for $J$-band data.  The
similarity of this figure to Figure \ref{fig:Himp} indicates the 
robustness of the $MID$ statistics across wavelength regimes.}
\label{fig:Jimp}
\end{figure}

In Figure \ref{fig:Jimp}, we display the relative importance of the
$CAS$-$GM_{20}$-$MID$ statistics for the detection of non-regulars (x-axis)
and mergers (y-axis) when we analyze $J$-band rather than $H$-band images.
(For these data, the smoothing scales were 0.75 and 1.4 pixels for the
non-regular and merger analyses, respectively.)
The conclusions that we draw from this figure and from Table \ref{tab:J} are
similar to those drawn from Figure \ref{fig:Himp} and Tables \ref{tab:HRNR}
and \ref{tab:HNMM}:
the $MID$ statistics are robust against changes in observation wavelength.

\begin{table}
\caption{Classifier Performance\,$\times$\,10$^4$\,$-$\,$J$-band}
\begin{tabular}{ccc}
\hline
 & ~~~Regular/Non-Regular~~ & ~~Non-Merger/Merger~~ \\
\hline
sens   & 7747 $\pm$ 14 & 7594 $\pm$ 18\\
spec   & 8124 $\pm$ 11 & 8295 $\pm$ 9\\
risk   & 4129 $\pm$ 8  & 4112 $\pm$ 14\\
toterr & 1973 $\pm$ 6  & 1770 $\pm$ 7\\
PPV    & 5947 $\pm$ 12 & 3143 $\pm$ 11\\
NPV    & 9119 $\pm$ 4  & 9715 $\pm$ 2\\
\hline
\label{tab:J}
\end{tabular}
This table displays sample mean 
plus/minus 1$\sigma$ standard error, 
each multipled by 10$^4$.
\end{table}

%
 
\subsection{Effect of changing the segmentation algorithm}

\label{subsect:segmap}

The {\tt SExtractor} segmentation maps used in the analysis of 
{\S}\ref{subsect:hband} associate image pixels with galaxies
using an absolute surface brightness threshold, such that the fraction of
galaxy flux within the map aperture varies with galaxy brightness.
This can introduce redshift-dependent biases into analyses due to surface
brightness dimming.
We verify that our results from {\S}\ref{subsect:hband} are robust to
segmentation algorithm by reanalyzing the data using an algorithm based on 
that of L04, who compute
a Petrosian radius for each galaxy, i.e., the radius at which the mean surface
brightness within an elliptical annulus is a fraction $\eta$ (e.g., 0.2)
of the mean brightness within that radius.  
The assumption of ellipticity will bias the construction of maps for
disturbed galaxies, so
we generalize the algorithm by using intensity quantiles.
We define a grid of quantile values and begin
with the largest value, determining which pixels have intensities greater than
this value and summing their intensities.  We then systematically
decrease the quantile value until the mean surface brightness of newly added
pixels is a fraction $\eta$ of the mean brightness of all pixels with 
intensities above the quantile value.  Because segmentation maps produced
by this algorithm are based on relative changes in surface brightness, these
maps are nominally redshift-independent.

For isolated, undisturbed galaxies that 
exhibit elliptical profiles, our algorithm yields maps similar
to those output by the L04 algorithm.  In other cases, when
distinct clumps of pixels are present, care must be taken since they
may represent two distinct
nuclei within one galaxy, or unrelated (i.e., non-interacting) pairs of 
galaxies, etc.  As we decrease the quantile value, we risk overblending,
but if we do not decrease the value enough, we risk missing clumps that
may be merger signatures.
Thus in our analysis we include a threshold quantile value below which
we do not blend distinct clumps of pixels (i.e., below which only one of the
observed clumps will be used to establish the segmentation map).  We 
determine the threshold value empirically by testing several values and 
finding which is associated with the smallest classification risk.

In Table \ref{tab:segmap}, we show classifier performance as a function of
algorithm and aperture parameter $\eta$.  We conclude that our new 
segmentation algorithm with $\eta \approx 0.2$ yields risk estimates on par 
with those yielded by the {\tt SExtractor} algorithm.  We observe that
classification degrades markedly with smaller apertures 
(i.e., larger $\eta$ values).  For larger apertures, classification
degrades more quickly for regular/non-regular analysis than for 
non-merger/merger analysis.  By examining other measures of classifier 
performance, we determine that this reduced ability to differentiate
between regular and irregular-but-not-peculiar galaxies is due more to
regulars being misclassified as irregulars than vice-versa.  This is 
consistent with the fact that smaller $\eta$ values will lead to increased
overblending and to machines classifying some fraction of the
regular galaxy population as non-regulars.

\begin{table}
\caption{Effect of $H$-band Segmentation: Estimated Risk\,$\times$\,10$^4$}
\begin{tabular}{ccccc}
\hline
Algorithm & Regular/Non-Regular & Non-Merger/Merger \\
\hline
{\tt SExtractor} & 3945 $\pm$ 8 & 3594 $\pm$ 13 \\
New ($\eta$ = 0.1) & 4229 $\pm$ 8 & 3643 $\pm$ 14 \\
New ($\eta$ = 0.2) & 3975 $\pm$ 8 & 3679 $\pm$ 13 \\
New ($\eta$ = 0.3) & 4081 $\pm$ 8 & 3960 $\pm$ 14 \\
New ($\eta$ = 0.4) & 4044 $\pm$ 8 & 3958 $\pm$ 12 \\
\hline
\label{tab:segmap}
\end{tabular}
This table displays sample mean 
plus/minus 1$\sigma$ standard error, 
each multipled by 10$^4$.
\end{table}

\subsection{Observation-specific effects}

\label{subsect:covar}

Previous works have shown that image statistics can be systematically affected 
by changes in observation-specific quantities like galaxy signal-to-noise
(e.g., $G$ and $M_{20}$, as shown in L04).  In this section,
we determine the effect of three observation-specific quantities on the
$MID$ statistics: galaxy signal-to-noise, galaxy size, and galaxy elongation.

We quantify a galaxy's signal-to-noise ($S/N$) by first
determining the sample mean $\bar{X}$ and sample standard deviation
$s_X$ of the intensities $I_{ij}$ of 
non-segmentation map pixels that lie within the galaxy's 
postage stamp.  We then standardize the intensities of all pixels in the
postage stamp:
\begin{eqnarray}
\hat{\left(\frac{S}{N}\right)}_{ij} = \frac{I_{ij} - \bar{X}}{s_X} \nonumber \,.
\end{eqnarray}
We examine the standardized intensities of those pixels
within $0.5\hat{r}$ of the galaxy's center, where
$\hat{r}$ is our estimate of galaxy ``size" in 
arc-seconds: $\hat{r} =0.06\sqrt{n_{\rm seg}/\pi}$.
We summarize the resulting empirical distribution by 
selecting the median standardized intensity.
Galaxy elongation is $e = 1-b/a$, 
where $a$ and $b$ are estimated semi-major and semi-minor galaxy axes.

\begin{figure}
\includegraphics[width=84mm]{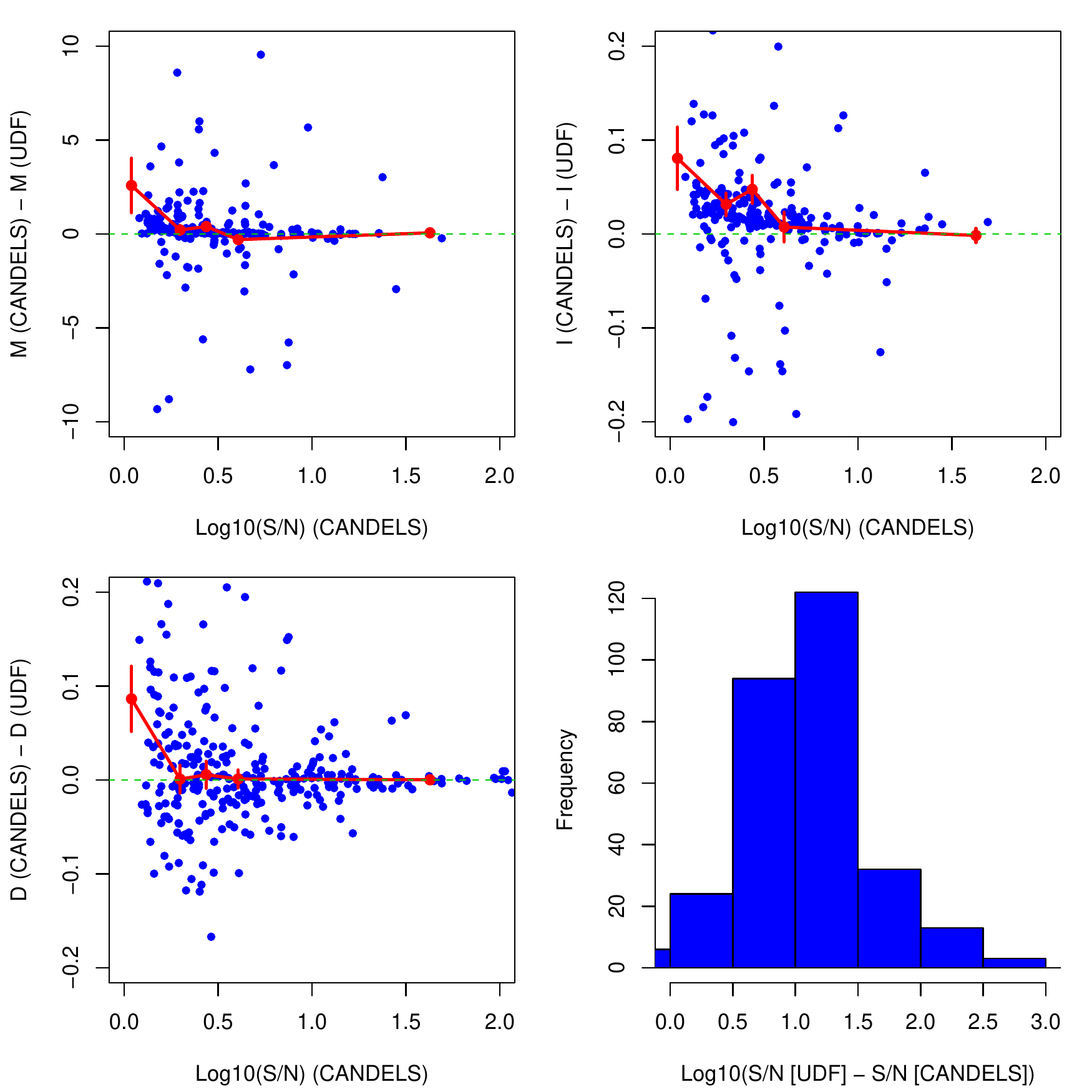}
\caption{
Estimated variation in the $MID$ statistics as a function of the 
signal-to-noise, $S/N$, between galaxies observed in a sample UDF field
($\approx$78.5 ks exposure) and the same galaxies observed in a 
$\approx$5.6 ks subexposure (commensurate with typical CANDELS exposure
times).  The blue dots are individual data and the red curves are estimates
of the mean created using five quantiles, i.e., the lowest 20\% of 
CANDELS-exposure $S/N$ values, the next 20\%, etc.  We find that
the $MID$ statistics are insensitive to $S/N$ in the regimes
$\gtrsim$ 1.7 ($M$ and $D$) and $\gtrsim$ 3.2 ($I$).
}
\label{fig:udfsn}
\end{figure}

\begin{figure}
\includegraphics[width=84mm]{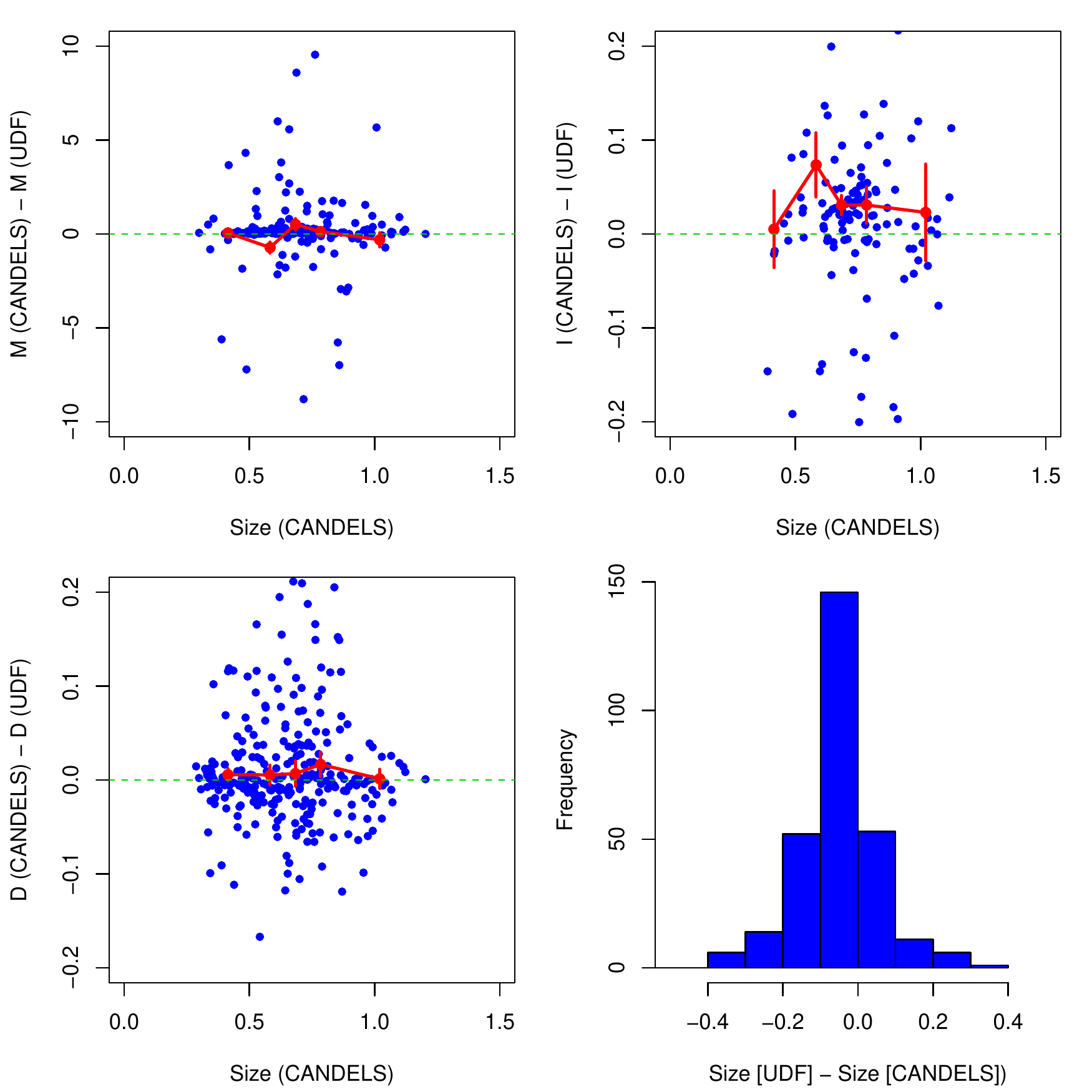}
\caption{
Same as Figure \ref{fig:udfsn}, except that estimated galaxy size 
in arcseconds is plotted along the $x$ axis.  We find that the $MID$
statistics are insensitive to galaxy size.  The data in the
upper right panel are consistent with the null hypothesis of a constant
offset from zero.  We find the amplitude of this offset is related to the 
scale of 
the smoothing kernel applied to the data prior to computing $I$.
See {\S}\ref{subsect:covar} for more detail.
}
\label{fig:udfsize}
\end{figure}

\begin{figure}
\includegraphics[width=84mm]{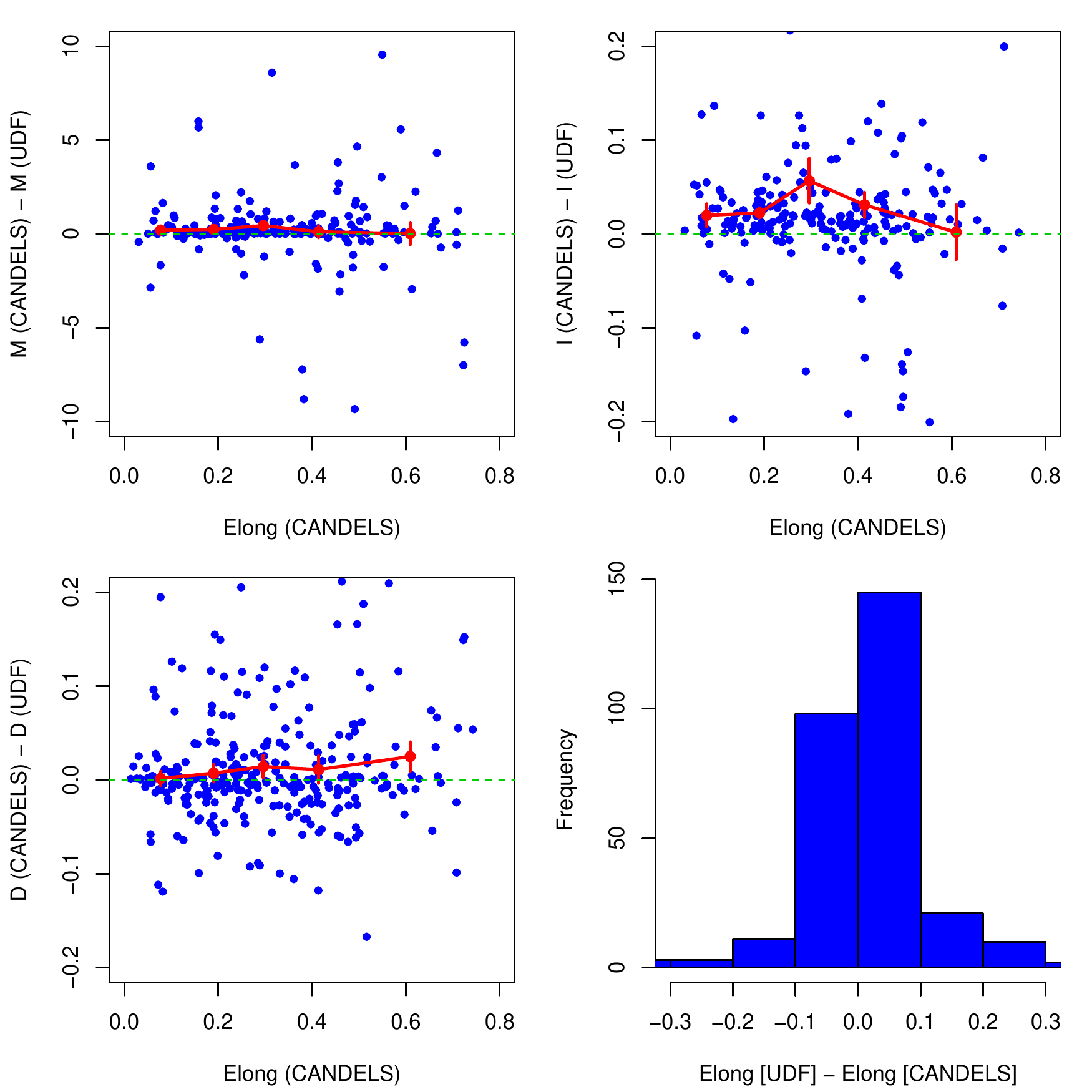}
\caption{
Same as Figure \ref{fig:udfsn}, except that estimated galaxy elongation
($1-b/a$) is plotted along the $x$ axis.  We find that the $MID$
statistics are insensitive to elongation, with the exception
of a systematic increase in $D$ in the regime $\gtrsim$ 0.48.  As in
Figure \ref{fig:udfsize}, we observe that the data in the
upper right panel are consistent with the null hypothesis of a constant
offset from zero and that the amplitude of this offset is related to the
smoothing kernel applied to the data prior to computing $I$.
See {\S}\ref{subsect:covar} for more detail.
}
\label{fig:udfelong}
\end{figure}

To examine how the $MID$ statistics vary as a function of $S/N$, etc., we 
follow the strategy of \cite{Lotz06} (see specifically Figure 1 and associated
text).  We analyze two images: an
$\approx$78.5 ks $H$-band WFC3 image of the Ultra Deep Field (UDF) and 
an $\approx$5.6 ks subset of that image, with the time of the shallower 
dataset chosen to be commensurate with typical exposure times of
CANDELS fields.  (The ERS data that we analyze in {\S}\ref{subsect:hband}
has integration time $\approx$50 ks.)  In Figures \ref{fig:udfsn}, 
\ref{fig:udfsize},
and \ref{fig:udfelong} we estimate how $MID$ statistic values
change with reduced exposure time. 
In these figures, each blue dot represents the value of
an observational quantity (plotted along the x-axis) and a change in 
statistic value (plotted along the y-axis).  We estimate the 
mean change in each statistic (the
red curves) by computing the 5\% trimmed sample mean and sample standard error
of those $y$ values associated with each of five quantiles along
the x-axis.  (The first quantile contains the first 20\% of the data, as
defined along the x-axis, the second contains the next 20\%, etc.)
We apply trimming so that our estimates are resistant to outliers.

On the basis of these figures, we conclude that the $M$ and $D$
statistics do not vary in any systematic fashion with $S/N$, size, or 
elongation, aside from the regime $S/N \lesssim 1.7$ for both and 
elongation $\gtrsim 0.48$ for $D$.  
(For the case of large elongation, where the systematic trend is not
necessarily obvious to the eye,
we test the null hypothesis of zero slope twice,
including and then discarding
the uppermost quantile of elongation values; 
the $p$ values are 0.021 and 0.221, respectively.
Thus if we include the uppermost quantile, we would conclude that the slope is
nonzero and thus that $D$ exhibits a systematic trend with elongation.)
We observe similar behavior for the 
$I$ statistic as a function of $S/N$, except that unlike $M$ and $D$
it does exhibit systematic changes in the regime $1.7 \lesssim S/N \lesssim 3.2$.
We also observe that as a function of size and elongation, $\Delta I$ exhibits 
an offset from zero that is consistent with being a constant offset (i.e.,
consistent with having slope zero) as determined via
weighted linear regression.  We find that this offset is sensitive to the
scale $\sigma$ of the bivariate Gaussian kernel that we use to smooth the
data prior to the computation of the $I$ statistic (see {\S}\ref{sect:istat}).  
In our analyses, we kept $\sigma$ constant between the longer and shorter
exposures.  However, 
a noisier (i.e., low exposure time) image requires a larger $\sigma$ to 
eliminate spurious secondary maxima.  We choose not to optimize the 
$\sigma$ values
in this analysis because doing so would not change the qualitative
conclusion that $I$ exhibits no systematic trends with either size and 
elongation.

\subsection{Effect of galaxy redshift}

\label{subsect:photoz}

\begin{figure}
\includegraphics[width=84mm]{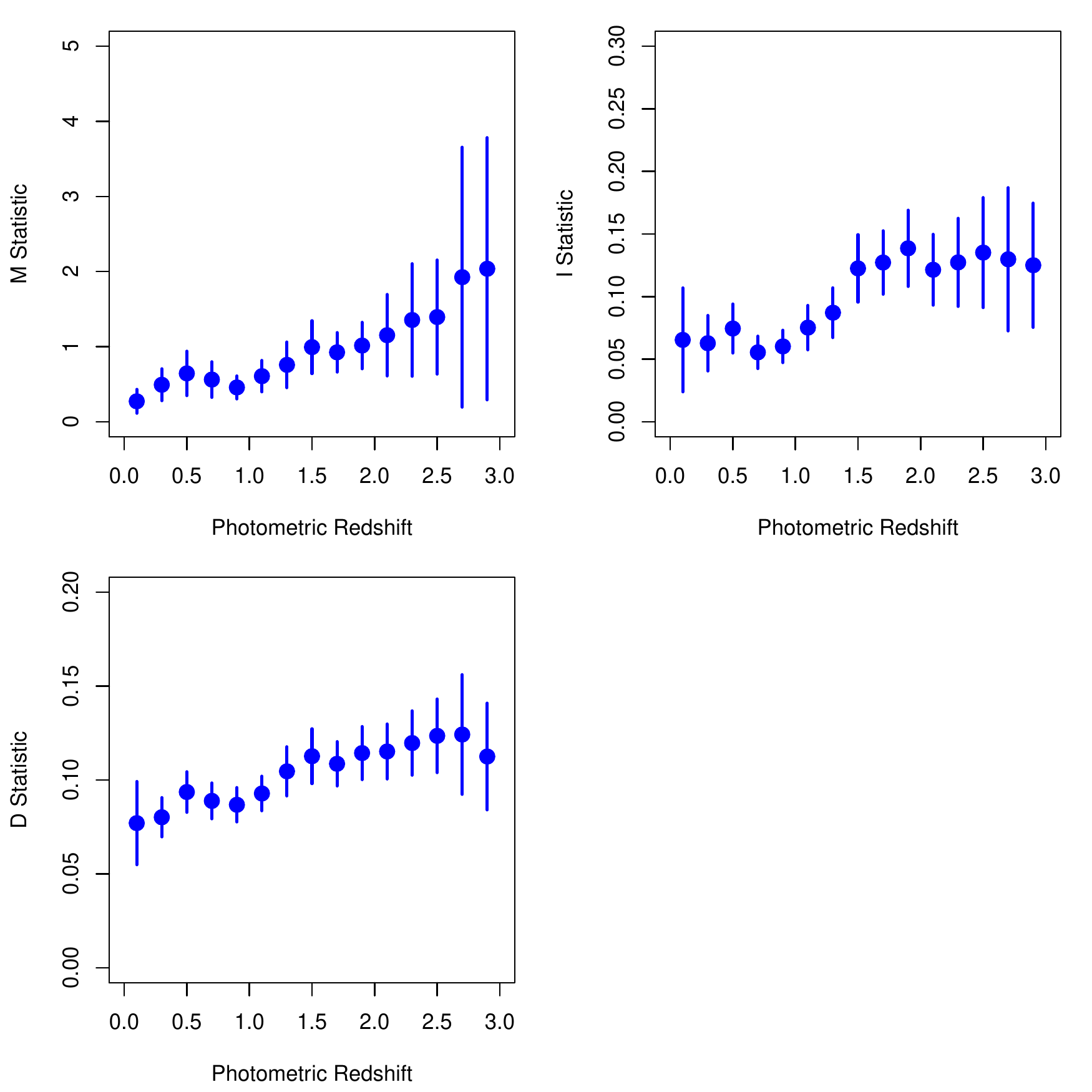}
\caption{
Means and standard errors of the means for the $M$, $I$, and
$D$ statistics as a function of photometric redshift in bins of size
$\Delta z = 0.2$, where the redshifts are provided by the GOODS-S catalog of
Dahlen et al.~(2010).  Via weighted linear regression, we find that the
$p$ values of the slopes are 6 $\times$ 10$^{-8}$ ($M$),         
2 $\times$ 10$^{-5}$ ($I$), and 10$^{-5}$ ($D$), i.e., we observe strong
positive correlations between redshift and the $MID$ statistics.
For more details, see {\S}\ref{subsect:photoz}.
}
\label{fig:midz}
\end{figure}

Having established the regimes in which the
$MID$ statistics are (in)sensitive to galaxy $S/N$, size, and elongation,
we verify that the ensemble average of the $MID$ statistics increases with 
redshift, as would expected for statistics that are sensitive to merging
activity.  In Figure \ref{fig:midz}, we show the means of the $M$, $I$, and
$D$ statistics, as well as their standard errors, as a function of 
photometric redshift in bins of size $\Delta z = 0.2$.
To construct this figure, we apply the GOODS-S
photometric catalog of \cite{Dahlen10}, which provides $z$ as well as
the values $z_{lo}$ and $z_{hi}$ that bound the central 95\% of each galaxy's
redshift probability density function (pdf).  Lacking further information,
we assume the pdf for galaxy $i$ to be a normal pdf with 
mean $\mu_i = (z_{i,hi}+z_{i,lo})/2$ and standard
deviation $\sigma_i = (z_{i,hi}-z_{i,lo})/3.92$.
Then, e.g., the estimated mean of $M$ in redshift bin $j$ is given by
\begin{eqnarray}
\bar{M}(z_j) = \frac{\sum_{i=1}^n w_{ij} M_i}{\sum_{i=1}^n w_{ij}} \,, \nonumber
\end{eqnarray}
where 
\begin{eqnarray}
w_{ij} = \Phi\left(\frac{(z_j+0.1)-\mu_i}{\sigma_i}\right) - \Phi\left(\frac{(z_j-0.1)-\mu_i}{\sigma_i}\right) \,. \nonumber
\end{eqnarray}
$\Phi(\cdot)$ is the cumulative distribution
function for the standard normal distribution.
We account for demonstrated biases by excluding
the six galaxies for which $S/N < 1.7$ in the upper left panel ($M$);
the twelve galaxies for which $S/N < 3.2$ in the upper right panel ($I$);
and the 297 galaxies for which $S/N < 1.7$ or $e > 0.48$ in the
lower left panel ($D$).

Figure \ref{fig:midz} shows clear trends between redshift and each of the
$MID$ statistics, but we should be careful when quantifying and interpreting
these trends because of our assumption of normal pdfs.  Thus here we 
simply assess whether the data are consistent with the null
hypothesis of no redshift trend (i.e., zero slope) via weighted linear 
regression.  The $p$ values of the slopes are 8.4 $\times$ 10$^{-8}$ ($M$), 
1.1 $\times$ 10$^{-5}$ ($I$), and 1.2 $\times$ 10$^{-6}$ ($D$);
we conclude that there are, as we would expect, strong
positive correlations between redshift and the $MID$ statistics.

\section{Examining the annotator's role}

\label{sect:meta}

In this section, we explore some of the effects that human annotators have 
on galaxy morphology analysis.  First, we ask whether 
annotators can accurately label mergers across the regimes of galaxy
$S/N$ and size spanned by
a single exposure, such as the $H$-band ERS data we analyze above.
Then we look towards the future and 
discuss two important issues: Is it always better to have
more annotators looking at each galaxy image? and 
Ultimately, are annotators even necessary within the
context of what we want to achieve, namely, selecting a model of hierarchical
structure formation and constraining its parameters?

\subsection{Can annotators accurately detect merging activity within a single dataset?}

\label{subsect:hilo}

\begin{figure}
\includegraphics[width=84mm]{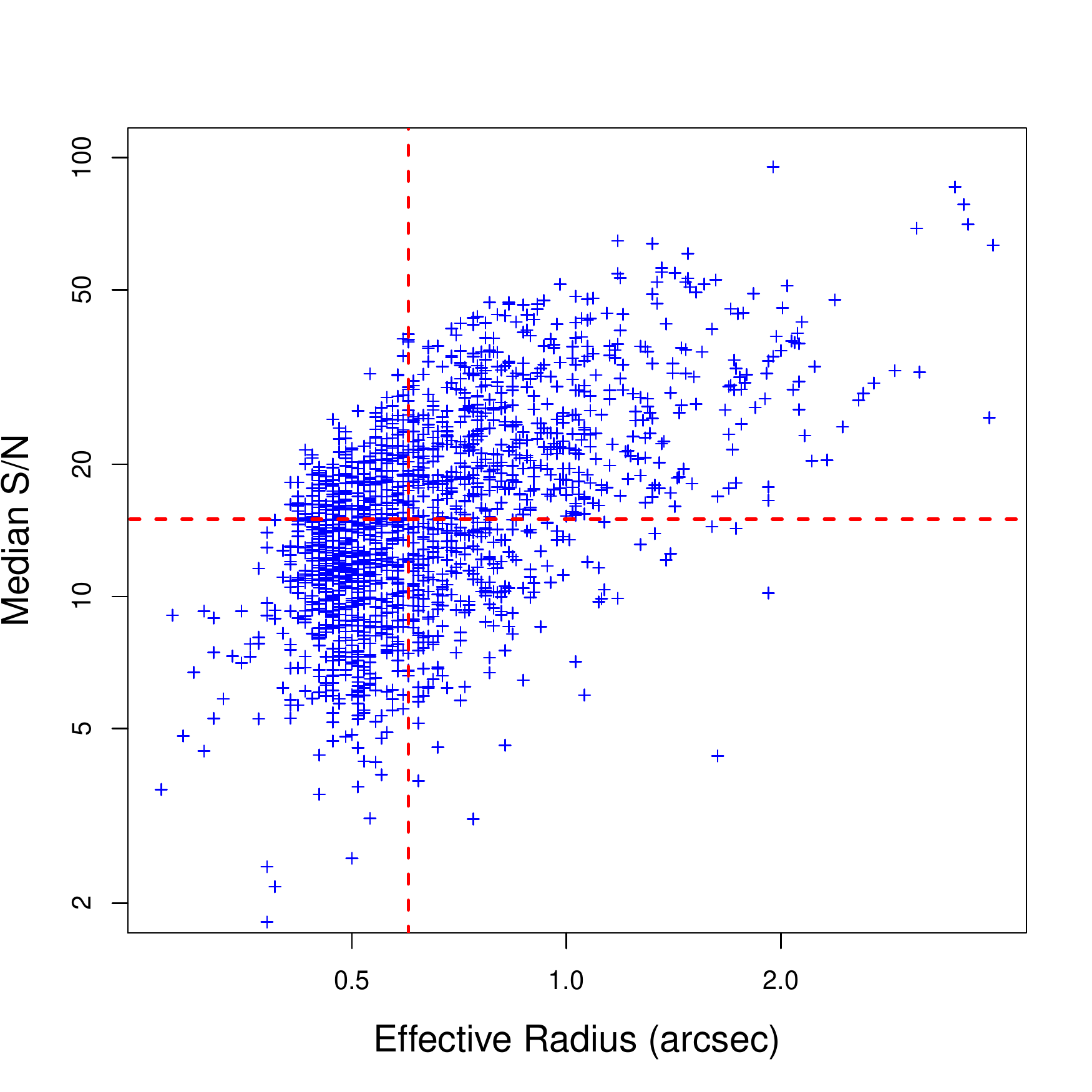}
\caption{Distribution of effective radius 
($\hat{r}$ in text) versus median pixelwise $S/N$
($\hat{\left(\frac{S}{N}\right)}$ in text) for our 1639-galaxy sample.
Dashed lines indicate the values
$\hat{r}$ = 0.6$\prime\prime$ and $\hat{\left(\frac{S}{N}\right)}$ = 15.
The 563 galaxies each within the lower-left and upper-right regions 
(as defined relative to where the dashed lines cross) comprise 
the ``low" and ``high" datasets respectively.
}
\label{fig:hilo}
\end{figure}

\begin{figure}
  \begin{minipage}[ht]{0.5\linewidth}
  \begin{center}
    \includegraphics[width=42mm]{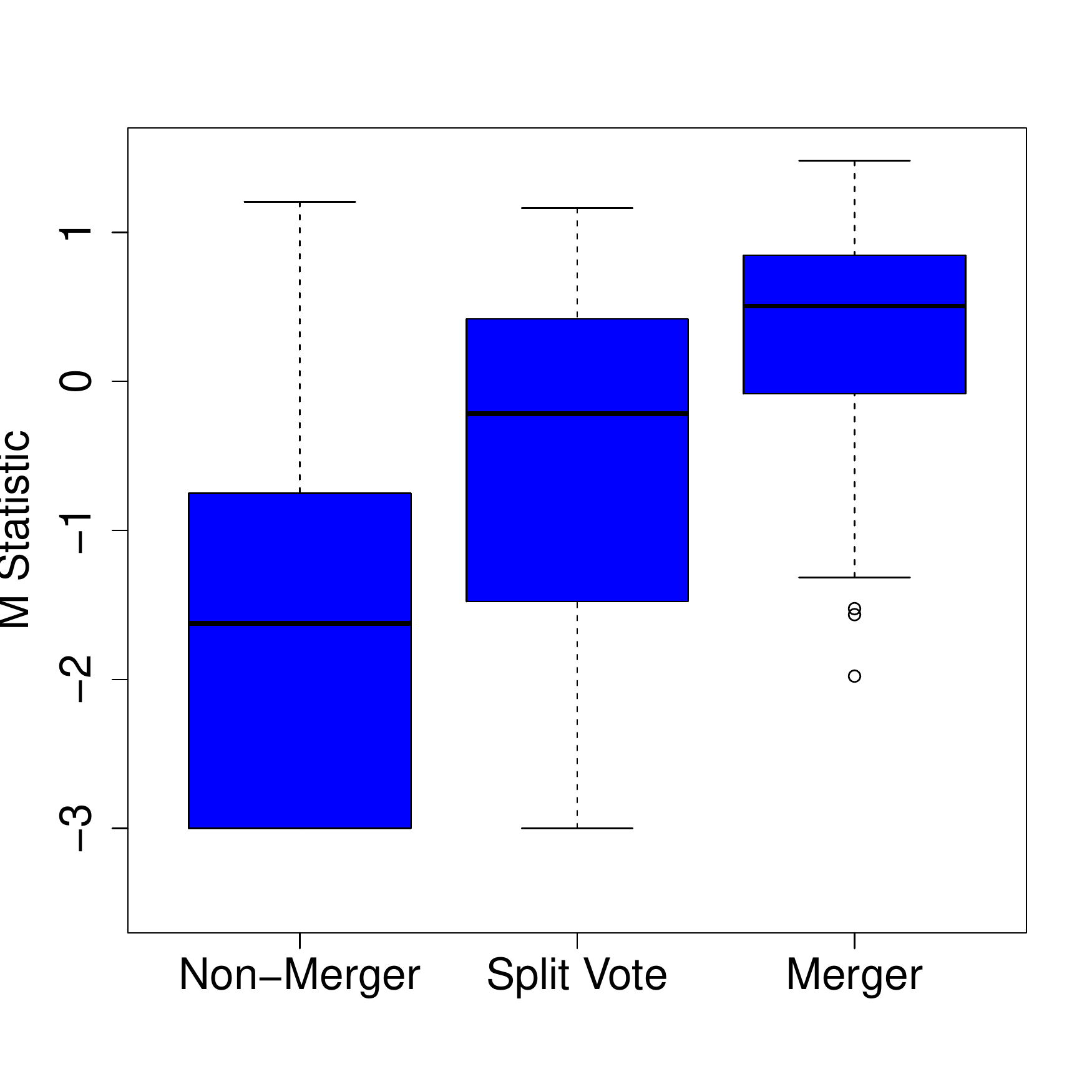}
  \end{center}
  \end{minipage}\hfill
  \begin{minipage}[ht]{0.5\linewidth}
  \begin{center}
    \includegraphics[width=42mm]{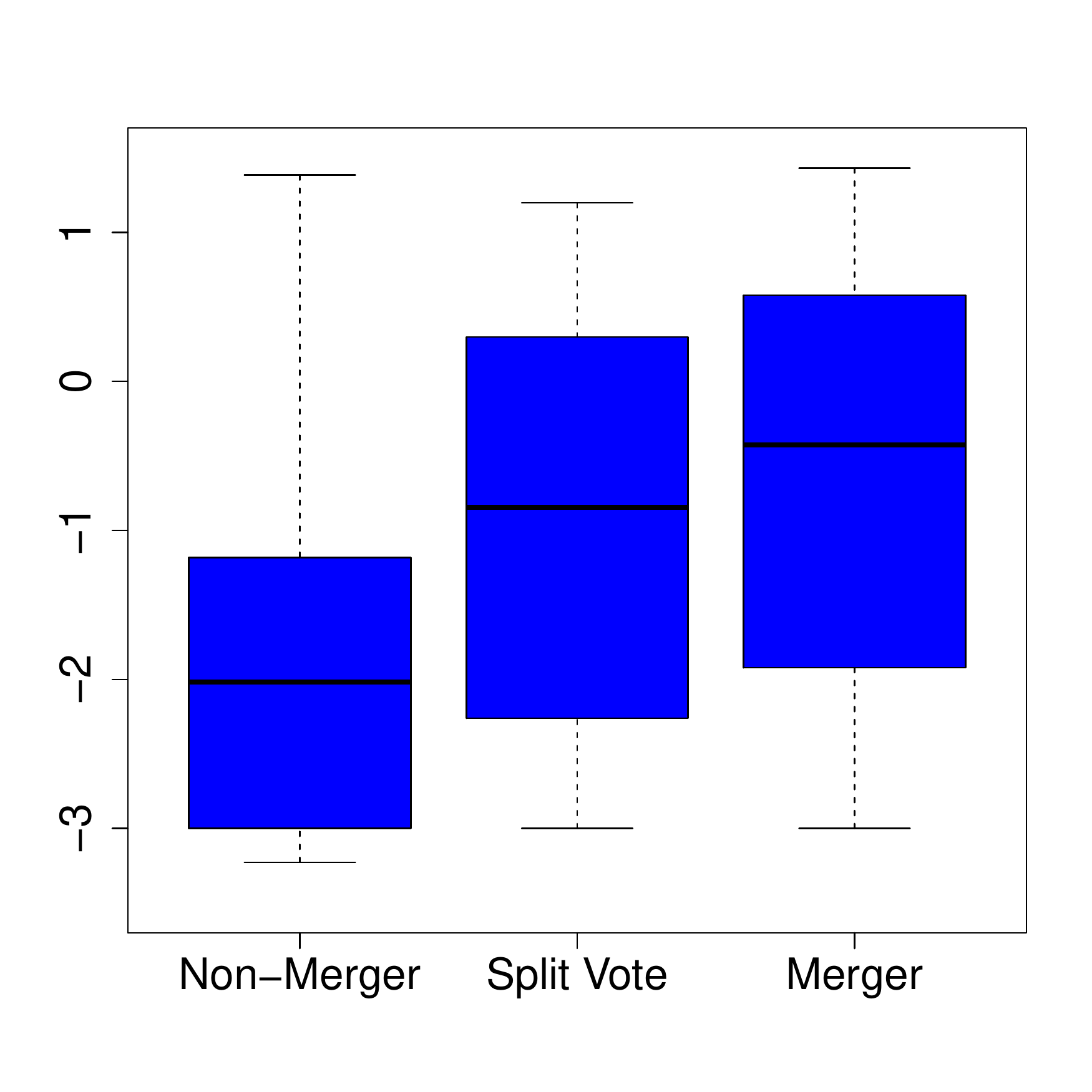}
  \end{center}
  \end{minipage}\hfill
\caption{Boxplots showing the distribution of $M$ statistic for
identified non-mergers ($f < 0.5$), galaxies for which the vote was split
($f = 0.5$), and mergers ($f > 0.5$).  The left and right panels show 
distributions for the ``low" and ``high" datasets respectively.  The
distributions are similar, except for the lack of identified mergers with
small $M$ values in the ``low" dataset: $M$ clearly correlates with the
ability of annotators to identify small-size-and-low-$S/N$ mergers.
The behavior of the $I$ statistic between datasets is similar and is not
shown.
}
\label{fig:boxplot}
\end{figure}

In {\S}\ref{subsect:hband}, we establish that the $MID$ statistics are useful
for detecting non-regular galaxies and merging galaxies that were labeled as
such by human annotators.  However, we have yet to establish whether labeling
is consistent as a function of, e.g., galaxy $S/N$ and size.
We construct two equally sized subsets ($n$ = 563)
of the 1639 galaxies in our data sample, one
where size and $S/N$ are both (relatively) low, and another where
size and $S/N$ are both (relatively) high:
\begin{eqnarray}
\hat{r} \leq 0.6'' ~~{\rm and}~~ \hat{\left(\frac{S}{N}\right)} \leq 15 ~~ && ~~ {\rm ``low"} \nonumber \\
\hat{r} > 0.6'' ~~{\rm and}~~ \hat{\left(\frac{S}{N}\right)} > 15 ~~ && ~~ {\rm ``high"} \nonumber
\end{eqnarray}
Both the ``low" and ``high" datasets contain 563 galaxies.  See Figure
\ref{fig:hilo}.  
In Figure \ref{fig:boxplot} we display
the distributions of the $M$ statistics for both datasets, for merger
vote fractions $f < 0.5$, $f = 0.5$, and $f > 0.5$.  We immediately
observe a lack of identified mergers ($f > 0.5$) with
small $M$ values in the ``low" dataset.  (We note that a similar issue
arises in the analysis of regular galaxies, as well as when we use the
$I$ statistic in place of $M$.)
{\em It is clear from Figure \ref{fig:boxplot} that
numerous small-$S/N$/size mergers are being mislabeled,
a systematic error that throws into doubt the idea that one 
can accurately estimate merger fractions at high redshifts via visual labeling.}
(Note that we base this conclusion on the analysis of a $\approx$50 ks image;
typical CANDELS exposures will be one-tenth as long,
exacerbating this error.)
This result helps motivate the alternative analysis paradigm that we discuss in
{\S}\ref{sect:meta_less}.

\subsection{The relationship between the number of annotators and
classification efficiency}

\label{sect:meta_more}

An astronomer's time is a valuble commodity.  Given a set of $N$ astronomers
with nominally similar annotation ability, is it best to have all of them
train a machine learning algorithm by visually inspecting hundreds if not
thousands of galaxy images?  Or can using a subset
of size $n \ll N$ yield similar detection efficiency?

To attempt to answer these questions, we utilize an analysis
carried out by the CANDELS collaboration (Kartaltepe et al.~2012, 
in preparation) in which 200 objects observed in the
$J$- and $H$-bands by the {\it HST} WFC3 in the DEEP-JH region of the
GOODS-S field
were each annotated by 42 voters.
The details of voting are similar to those
described above in {\S}\ref{sect:apply}, except that in this analysis each
annotator's vote is recorded, so there is no ambiguity about
the fraction of annotators identifying particular galaxies as either mergers or 
non-regulars.

After removing 15 objects from the sample that were subsequently identified
as stars (J.~Lotz, private communication), we analyze the 
$H$-band images of the remaining 185 galaxies
in a manner similar to that described in {\S}\ref{sect:apply}. 
The principal difference between analyses is that for computational
efficiency, we use only the $MID$ statistics; it is not imperative to
use all available statistics because our aim in this analysis is to observe how
the estimated risk varies as a function of the number of annotators, $n$,
without regard to its actual value.  

We assume that a new expert voter will randomly identify a given galaxy $i$ as 
a non-regular/merger with probability $p_i$, where $p_i$ is the recorded
vote fraction for the set of 42 annotators.  Thus to simulate the number of 
votes for non-regularity/merging for each galaxy, given $n$ annotators, we
sample from a binomial distribution with parameters $n$ and $p_i$.  The result
is an integer number of votes $Z_i \in [0,n]$, with the simulated vote
fraction being $F_i = Z_i/n$.  Given $F$ and the $MID$ statistics for all
185 galaxies, we run random forest and output the estimated risk.  
We repeat the process of simulation and risk estimation 100 times for each
value of $n$ so as to build up an empirical distribution of estimated
risk values.  Note that as we increase $n$, we only randomly sample {\em new}
votes.  For instance, to go from $n$ = 5 to $n$ = 7, we add two new simulated
votes to the five we already have.  We feel that this is
more realistic than randomly sampling a completely new set of votes,
as an increased $n$ in practice generally will be implemented by adding
to a core group of annotators rather than replacing that core group in its
entirety.

\begin{figure}
  \begin{minipage}[ht]{0.5\linewidth}
  \begin{center}
    \includegraphics[width=42mm]{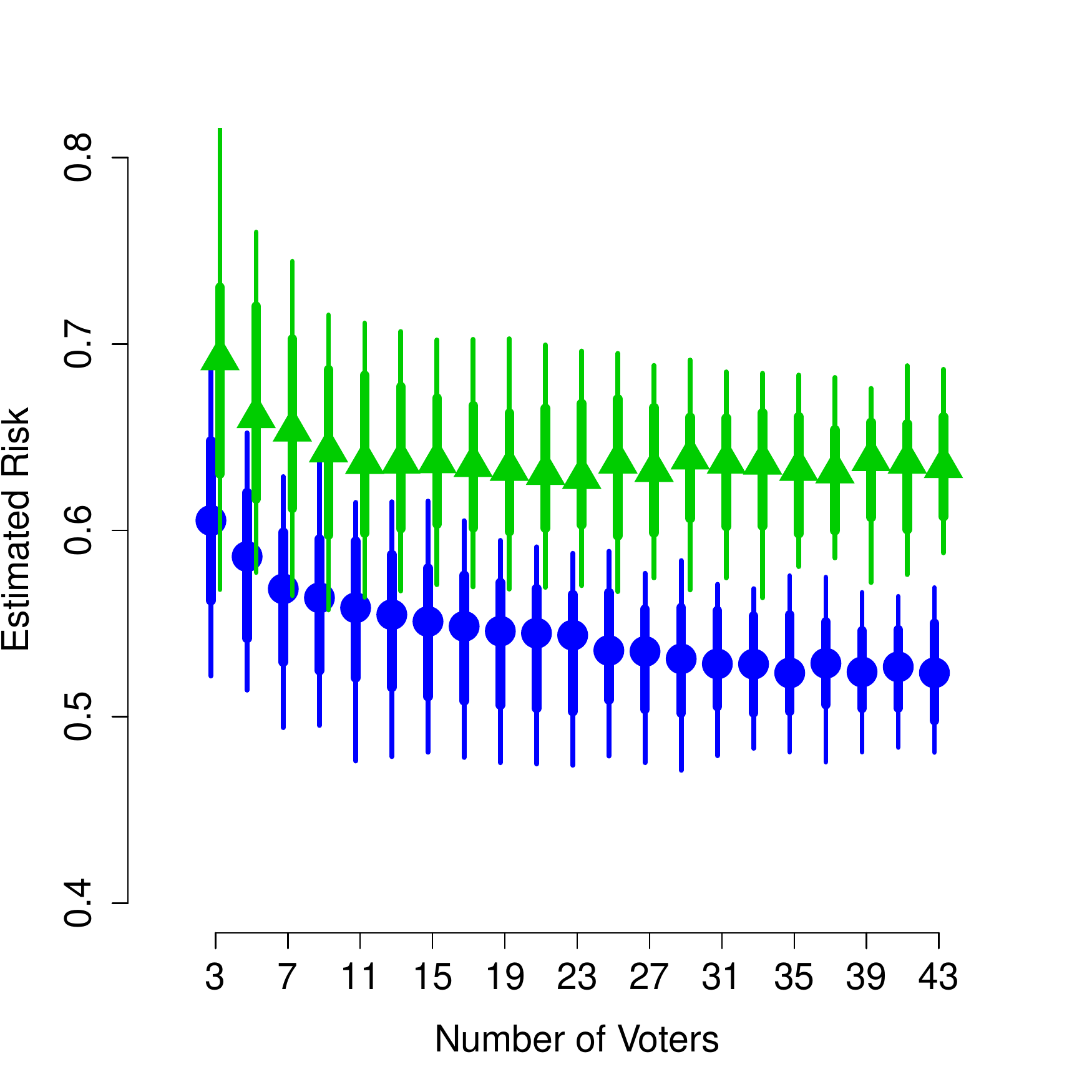}
  \end{center}
  \end{minipage}\hfill
  \begin{minipage}[ht]{0.5\linewidth}
  \begin{center}
    \includegraphics[width=42mm]{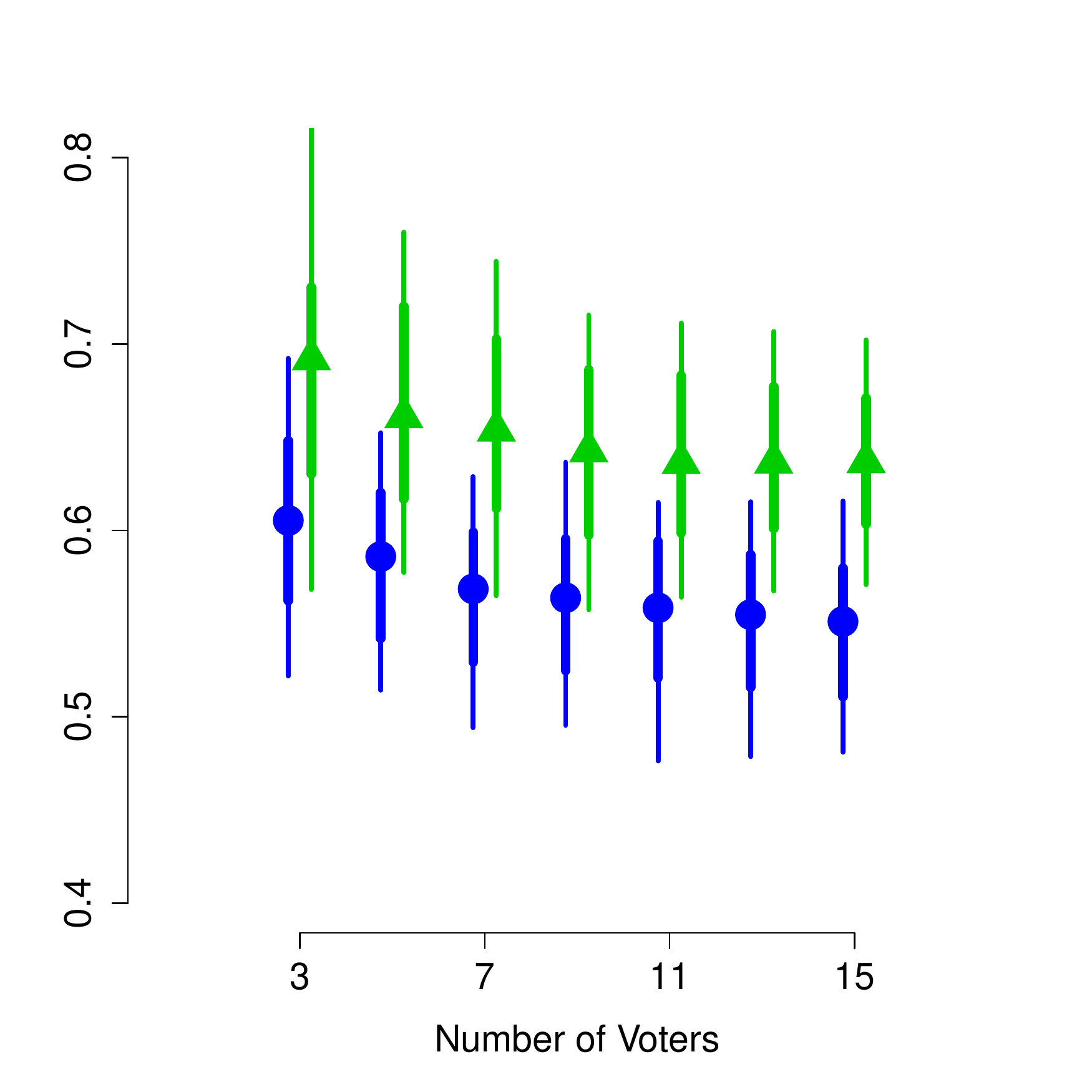}
  \end{center}
  \end{minipage}\hfill
\caption{Median estimated risk in the detection of non-regulars 
(the lower sequence of points in each panel, denoted with blue circles)
and mergers 
(the upper sequence of points in each panel, denoted with green triangles)
in the analysis of 185 galaxies described in 
{\S}\ref{sect:meta_more}, as a function of number of annotators.
The right panel shows the same data as the left, for the reduced range
$n$ = 3-15.  The thin and thick lines drawn through each point represents the 
central 95\% and 68\% of the empirical distribution of risk 
values, respectively.  The lines are slightly offset from each other for
clarity.  See {\S}\ref{sect:meta_more} for further discussion.
}
\label{fig:exrisk}
\end{figure}

In Figure \ref{fig:exrisk} we display
the median of our risk distributions for
the non-regular (blue points and lines) and merger (green points and lines) 
detection cases.
The thin and thick lines drawn through each point
indicate the range for the central 95 and 68 values in each distribution,
respectively.
(Note that the values of the risk are generally much higher here than in
Tables \ref{tab:HRNR} and \ref{tab:HNMM} because the training sets here
are one-ninth the size of those in the analysis of {\S}\ref{sect:apply}.)
We observe that for both cases, the estimated risk decreases somewhat
sharply when $n \lesssim$ 10; above $n \approx$ 10,
the risk for the non-regular case still decreases, albeit
more slowly, while the risk for the merger case remains constant.
Imprecise estimation of the true vote fractions for small $n$ and
vote fraction discretization lead to the 
increase in risk as $n \rightarrow$ 0, as
it becomes less and less likely that, e.g., a ``true" merger will be identified
as a merger by both annotators {\em and} the machine classifier.

\begin{figure}
\includegraphics[width=84mm]{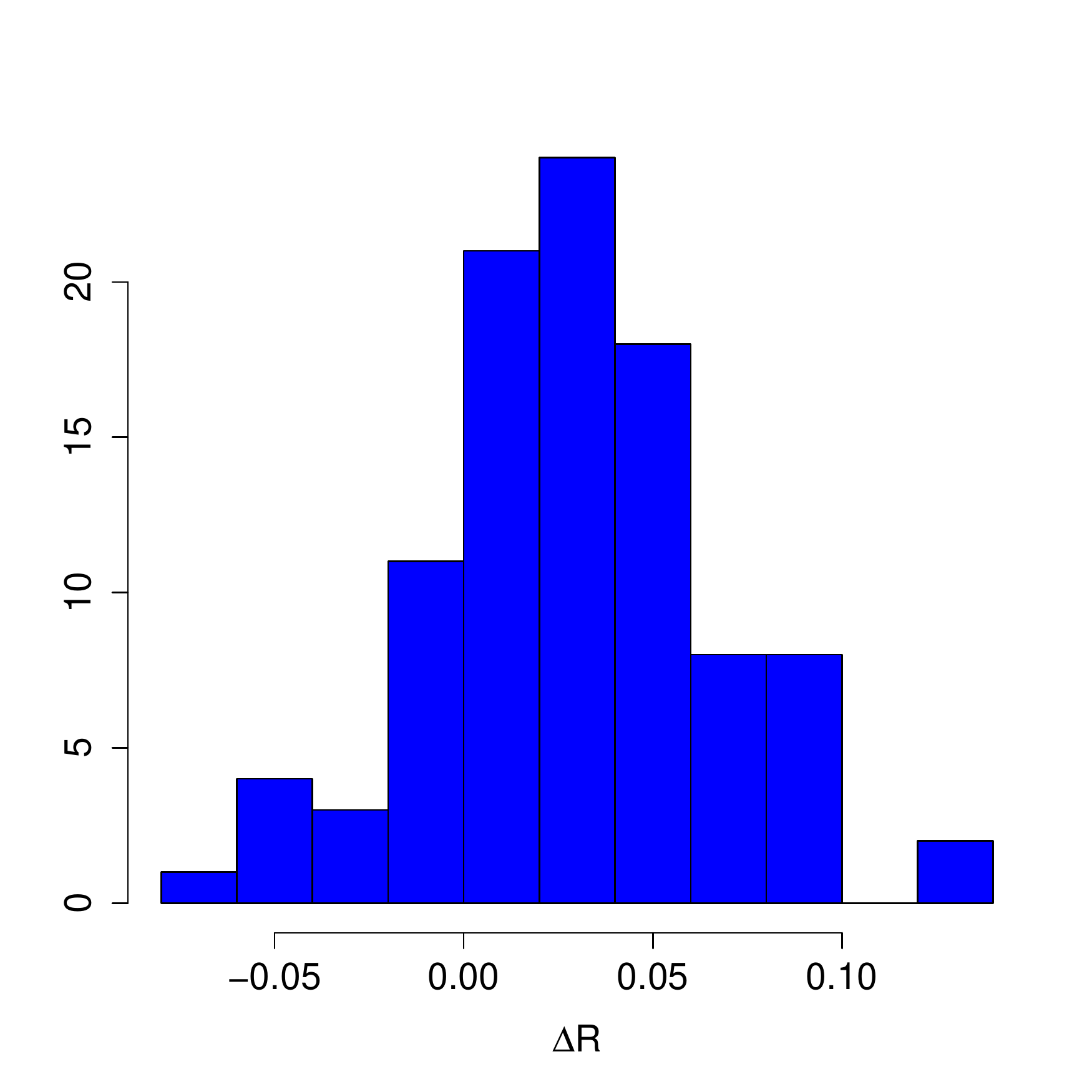}
\caption{Histogram of the changes in estimated risk that occur when we
increase the number of annotators from 11 to 43 in each of the 100 
simulations we run in our analysis of 185 galaxies ({\S}\ref{sect:meta_more}).
Positive values of $\Delta R$ indicate a reduction in estimated risk.  In 
19 of 100 simulations, the estimated risk increases: adding annotators led
to worse results.
}
\label{fig:delrisk}
\end{figure}

For the merger case, it is clear from Figure \ref{fig:exrisk} that
little improvement in risk estimation occurs when adding annotators beyond
$n \approx 10$.  For the non-regular case, there is a slight improvement
{\em on average}, but there is no guarantee that one would see that 
improvement in any {\em single} analysis.  In Figure \ref{fig:delrisk} we show
the histogram of the change in risk, $\Delta R$, 
that occurs as we go from $n =$ 11 to $n =$ 43 annotators; each value is
derived from one of our 100 simulations.  In 19 of 100 cases, there is
an {\em increase} in estimated risk; adding 32 annotators made our results
worse.  This lack of significant improvement in estimated risk, coupled with
the time resources that would be expended by the additional annotators, 
argues strongly that an annotator pool of size $n \approx$ 10 is sufficient
for detecting non-regulars.

{\em We conclude
that no more than $\approx$10 annotators are needed to effectively train a
galaxy morphology classifier using a given set of galaxy images 
when the goal is to detect non-regular galaxies or mergers.}  If more
annotators are available, they should examine additional sets of galaxy
images to increase the overall training set size, and thereby reduce the
misclassification risk.

\subsection{Towards the future: eliminating visual annotation}

\label{sect:meta_less}

As hinted at throughout this work, there are many issues with
annotating galaxies and using the resulting morphologies to make quantitative
statements about structure formation.  Some of the more noteworthy issues
are the following:
\begin{itemize}
\item {\it Ambiguity.}  Expert annotators often do not agree on whether a 
given galaxy is, e.g., undergoing a merger (as opposed to, e.g.,
undergoing star formation due to {\it in situ} disk instabilites).  
This inability to agree, which led to the large
spread of merger fraction estimates compiled by \cite{Lotz11}, is a 
not-easily quantified source of systematic error: e.g., how does one
incorporate the experience and innate biases of each annotator into a
statistical analysis?
Given the subject of this paper, ambiguity is perhaps an obvious issue to 
point out, but its deleterious effects on structure formation analysis 
cannot be overstated.  
\item {\it Loss of Statistical Information.}  Above and beyond the 
issue of ambiguity is
the fact that in the classification exercise, we are attempting to take a
continuous distribution (e.g., all possible galaxy morphologies)
and discretize it (reduce it to, e.g., two bins: mergers and non-mergers).
Discretization can only have an adverse effect on statistical inferences,
making them certainly less precise and perhaps less accurate.
\item {\it Waste of Resources.}  Annotation is, by definition, a time-consuming
exercise that diverts astronomers from other activities.
\end{itemize}

Our vision of (near) future analyses of galaxy morphology and hierarchical 
structure formation rests on the belief that simulation engines will be
developed that can replicate the wide variety of observed morphologies at a
resolution at least on par with current observations.  
If this occurs, then we can fit structure formation models in the following
manner:
\begin{enumerate}
\item Populate space of image statistics by analyzing a set of observed galaxies.
\item Pick a set of model parameters describing structure formation.
\item Run a simulator and project the simulated galaxies down onto 
a (set of) two-dimensional plane(s).
\item Populate a space of image statistics 
by analyzing the set of simulated galaxy images.
\item Directly compare the estimated distributions of the simulated and 
observed statistics.
\item Return to step (ii), changing the model parameter values, and iterate
until convergence is achieved.
\end{enumerate}
The comparison step, step (v), involves estimating the density functions from
which the simulated and observed statistics were sampled, and then determining
a ``distance" between those functions.  There exist numerous, mature
methodologies for performing density estimation and estimating distances
between density functions.  (A summary of possible distance 
measures is provided in, e.g., \citealt{Cha07}.)  The key to a computationally
efficient comparison is to avoid the ``curse of dimensionality": density
estimation is difficult in more than even a few dimensions.  Thus even if
annotators are no longer needed, 
{\em there will always be a need
to define new statistics that can better disambiguate the
morphologies of galaxies.}

\section{Summary}

\label{sect:summary}

This work is motivated by the problem of detecting irregular and
peculiar galaxies in an automatic fashion in low-resolution and 
low-S/N images.  This information, combined with estimates of galaxy redshift,
can help us determine how the merger fraction evolves with time and
thus place constraints on theories of hierarchical structure formation.
One body of work on irregular/peculiar galaxy detection 
focuses on the tactic of
reducing galaxy images to a set of summary statistics that
sufficiently captures morphological details and allows computationally
efficient analysis of large samples of galaxies.  In particular,
the concentration, asymmetry, and clumpiness ($CAS$) statistics of
C03 and references therein and the Gini and $M_{20}$
statistics of L04 and references therein have become standard
statistics to use in morphological analyses.
However, the utility of these statistics to detect the
irregularity or peculiarity of high-redshift galaxies is open
to question, with simulations by L04 in particular suggesting 
that the $GM_{20}$ statistics would lose the ability to detect peculiar
galaxies in the low-resolution/low-S/N regime.

In an attempt to increase detection efficiency, we have developed three
new image statistics$-$the multi-mode ($M$), intensity ($I$), and
deviation ($D$) statistics$-$and we test them (along with $CAS$ and $GM_{20}$)
on $J$- and $H$-band {\it HST} WFC3 images of 1,639 
galaxies in the GOODS-South field.
In particular, we test these statistics' abilities to identify both
irregular and peculiar galaxies (which we collectively dub ``non-regular") and
peculiar galaxies alone (or ``mergers").  
We use a machine learning-based classifier, random forest,
to predict the classes of each of 1,639 galaxies, and we determine its
performance by comparing these predictions to visual annotations made by
members of the CANDELS collaboration.
We strongly advocate the use of random forest or other, similar algorithms
such as support vector machines in galaxy morphology studies,
as they allow computationally
efficient analyses of high-dimensional image-statistic spaces,
and thus stand in contrast to the commonly used inefficient technique of
projecting these spaces to two-dimensional planes within which classes
are identified.\footnote{
We note that generally one cannot predict {\it a priori} which 
machine learning-based algorithm is best to use for a particular analysis,
so we also strongly advocate using more than one to ensure robust results.
For instance, in this work we tested four, and found all to give similar
results; random forest was subsequently chosen because of the four
it is conceptually the simplest.}

As shown in Figures \ref{fig:Himp} and \ref{fig:Jimp} and discussed in 
{\S}{\S}\ref{subsect:hband}-\ref{subsect:jband},
we find that our $MID$ statistics, along with the asymmetry statistic $A$,
are the most important ones for disambiguating sets of galaxies in our sample;
in general, using these four statistics alone yields detection 
efficiencies on par with using the full set $CAS$-$GM_{20}$-$MID$.
In {\S}\ref{subsect:segmap}, we demonstrate that classifier performance is
insensitive to the details of the algorithm for constructing segmentation 
maps, and in {\S}\ref{subsect:covar}, we find
that the $MID$ statistics are largely insensitive to changes in galaxy
signal-to-noise, size, and elongation.

We explore the role of human annotators in {\S}\ref{sect:meta}.
In {\S}\ref{subsect:hilo}, we construct two subsamples of our dataset,
with small-$S/N$/size and large-$S/N$/size galaxies, respectively,
to ascertain whether the ability of annotators to label mergers or
non-regulars degrades with $S/N$ and size.
The difference in appearance of the right-most boxplots in both panels of
Figure \ref{fig:boxplot} strongly suggests an inability on the part of
annotators to properly label mergers with relatively small values 
of $M$ in low-$S/N$/size data.  
(Similar results hold for regulars vs. non-regulars, and if we use
$I$ in place of $M$.)
Beyond any numbers, this result 
raises doubt about whether merger rates at high redshift can ever be
accurately estimated using annotators.

We next assess
how the number of annotators affects classification performance, using a set of
185 $H$-band-observed {\it HST} WFC3 images that were each annotated
by 42 members of the CANDELS collaboration.   We repeatedly
sampled subsets of these annotators and used their votes to generate new sets
of class predictions, and then we recorded how the estimated risk of making
an incorrect prediction varied as a function of the number of annotators
(see Figure \ref{fig:exrisk}).  As discussed in
{\S}\ref{sect:meta_more}, we find that there is no evidence that increasing
the number of annotators above $n \approx$ 10 yields any improvements in
classifier performance for a given set of galaxy images; if more are
available, they should be charged with increasing the training set size
by annotating additional galaxy images, thereby reducing the risk of
misclassification.

As we discuss in
{\S}\ref{sect:meta_less}, however, any argument over the optimal number of
annotators to deploy within a project may became moot in the future
if simulation engines are developed that can effectively recreate the
observed populations of galaxies.  In our vision of the future, 
an analyst would populate two
spaces$-$one with observed statistics, and one with statistics computed
from simulated galaxies$-$and estimate and directly compare the density 
functions from
which the statistics were sampled.  In other words, we would quantitatively
determine how well
the points in the two spaces ``line up."  This process would be repeated
until an optimal match is found, i.e., until the best-fit model of hierarchical
structure formation is found.  This methodology effectively sidesteps the issue
of, e.g., estimating the merger fraction as a function of redshift, 
but one could still determine that by, e.g., adopting a definition of
``merger" and examining the evolutionary histories of galaxies in the
best-fit simulation to see which underwent the process.  
While annotation is eliminated in this
vision, the need for new and improved image statistics is not,
since to avoid the ``curse of dimensionality" we would always strive to
perform density estimation in relatively low-dimensional spaces of image
statistics.

\section*{Acknowledgements}

The authors would like to thank the members of the CANDELS collaboration for
providing the proprietary data and annotations upon which this work is based,
and in particular would like to thank E.~Bell
for helpful comments.  The authors would also like to thank the anonymous
referee for helpful commentary.  This work was supported by NSF 
grant \#1106956.  RI thanks the
Conselho Nacional de Desenvolvimento Cient\'{i}fico e Tecnol\'{o}gico
for its support.
Support for {\em HST} Program GO-12060 was provided by NASA through grants from
the Space Telescope Science Institute, which is operated by the Association of
Universities for Research in Astronomy, Inc., under NASA contract NAS5-26555.

\appendix

\section{Implementing random forest using R}

\label{app:ranfor}

{\tt R}, an open source application for statistical computing available
at {\tt http://www.r-project.org}, is widely used in the statistics
community.  One of the primary benefits to using {\tt R} is that one does
not have to write code to implement commonly used statistics
and machine learning algorithms, which generally exist in one or more
packages contributed to the Comprehensive R Archive Network (CRAN).
One such package is {\tt randomForest}, which we utilize here.

After {\tt R} is downloaded and the GUI is opened, the first step is
to install the {\tt randomForest} package.  This may be done by, e.g.,
tying the following at the prompt within the GUI window and following
any subsequent directions:
\begin{verbatim}
> install.packages("randomForest")
\end{verbatim}

Before running random forest, however, it is good practice to create a
{\em source file}, which is a list of commands that can be read into {\tt R}
via the {\tt source} command (or via the GUI's pull-down menus).  In the
following, we assume that the image statistics and the vote fractions
are in ASCII text files with single-row headers and one additional row
for each galaxy, e.g.,
\begin{verbatim}
M         I        D
0.4747    0.3123   0.5666
0.0133    0.0405   0.0259
...
\end{verbatim}
We dub these files {\tt statistics.txt} and {\tt votes.txt}.

The following are the contents of a minimalist file
that when sourced will run random
forest over 1000 splits while assuming that half the galaxies are to
be assigned to the training set.
\begin{verbatim}
# Input the random forest library functions
library(randomForest)

# Input the dependent (votes) and 
#   independent (statistics) data
# Assume the first row of votes is:
#   "nonreg merger"
# Assume the first row of statistics is:
#   "M I D"
#
votes = read.table("votes.txt",header=T)$merger
stats = read.table("statistics.txt",header=T)

# Standardize the statistics column-by-column: 
#   X_i -> (X_i-mean(X))/sd(X)
#
stats = scale(stats)

# Specify the (sub)set of statistics to input to 
#   random forest.
set = c("M","I","D")

# Here we assume no ties that have to be broken.
class = votes>0.5

ntrain = round(0.50*length(data[,1]))
B = 1000

# Initialize vectors of length B
sens = spec = risk = toterr = ppv = npv = rep(-9,B)

for ( ii in 1:B ) {
  # assign galaxies to training/test sets
  train     = sample(1:length(stats[,1]),size=ntrain)
  test      = (1:length(stats[,1]))[-train]
  # run random forest regression -- Section 3.1
  fit       = randomForest(x=stats[train,set],
                           y=votes[train],
                           maxnodes=8)
  predTrain = predict(fit,stats[train,set])
  predTest  = predict(fit,stats[test,set])
  # determine the class threshold -- Section 3.2
  cut = seq(0,1,0.001)
  c0 = NULL
  c1 = NULL
  for ( ii in 1:length(cut) ) {
    c0 = append(c0,
         mean(predTrain[class[train]==0]>cut[ii]))
    c1 = append(c1,
         mean(predTrain[class[train]==1]<cut[ii]))
  }
  smooth = ksmooth(cut,colMeans(rbind(c0,c1)),
                   bandwidth=0.05)
  mc = smooth$x[which.min(smooth$y)]
  # record the important quantities -- Section 3.3
  sens[ii]   = 1-mean(predTest[class[test]==1]<mc)
  spec[ii]   = 1-mean(predTest[class[test]==0]>=mc)
  risk[ii]   = mean(predTest[class[test]==1]<mc)+
               mean(predTest[class[test]==0]>mc)
  toterr[ii] = mean((predTest>mc)!=class[test])
  ppv[ii]    = 1-mean(class[test][predTest>mc]==0)
  npv[ii]    = 1-mean(class[test][predTest<=mc]==1)
}

# Output the mean estimated risk over all B splits 
#   (other values can be output in a similar manner).
cat("Estimated Risk = ",mean(risk),"\n")

q()
\end{verbatim}

Once this file is saved to disk (we dub this file
{\tt rf\_source.R}), it may be sourced via the
``Source File..." option in the {\tt R} GUI's pull-down
menu, or by typing
\begin{verbatim}
> source("<path>/rf_source.R")
\end{verbatim}

\bsp

\label{lastpage}

\end{document}